%% file: ANA-TOPQ-2023-35-PAPER.tex
\pdfoutput=1
\documentclass[PAPER, atlasdraft=false, UKenglish, cernpreprint, texmf, orcidlogo]{atlasdoc}
\usepackage{atlaspackage}
\usepackage[utf8]{inputenc}
\DeclareUnicodeCharacter{202F}{~}
\usepackage{atlasbiblatex}
\usepackage{multirow}
\usepackage{makecell}

\usepackage{atlasphysics}
\usepackage{amsmath}

\graphicspath{{logos/}{figures/}}

\usepackage{ANA-TOPQ-2023-35-PAPER-defs}
\AtlasTitle{Search for charged-lepton-flavour violating $e\mu tq$ interactions at $\sqrt{s}=13$ TeV with the ATLAS detector}

\AtlasAbstract{%
A search for charged-lepton-flavour violation (cLFV) in top-quark production and decay is presented. The data analysed correspond to $140~\textrm{fb}^{-1}$ of proton--proton collisions recorded at a centre-of-mass energy of $\sqrt{s}= 13$ TeV by the ATLAS detector at the Large Hadron Collider.
The possible cLFV contribution is studied via two mechanisms. The first is via the production of a leptonically-decaying single-top quark in association with a pair of oppositely-charged different-flavour leptons. The second is the production of top-quark pairs, where one top quark decays into a pair of oppositely-charged different-flavour leptons and a $u$- or $c$-quark, while the other decays into a leptonically-decaying $W$ boson and a $b$-quark, according to the Standard Model. The signal processes contain three charged leptons, taken to be electrons or muons.
No significant deviation from the Standard Model prediction is observed, and these results are interpreted within an Effective Field Theory framework. Upper limits are placed at 95\% confidence level on six Wilson coefficients, ranging from $|C^{\mathrm{tensor}}_{e\mu tu}|/\Lambda^2 < 0.018~\mathrm{TeV}^{-2}$ to $|C^{\mathrm{scalar}}_{e\mu tc}|/\Lambda^2 < 0.32~\mathrm{TeV}^{-2}$ for the tensor $u$-quark and scalar $c$-quark couplings, respectively. These Wilson coefficient limits are translated into 95\% confidence level upper limits on cLFV top-quark decay branching ratios, ranging from $\mathcal{B}(t \to e\mu u) < 6 \cdot 10^{-9}$ to $\mathcal{B}(t \to e\mu c) < 2.7 \cdot 10^{-7}$ for the scalar $u$-quark and tensor $c$-quark couplings, respectively.
}

\AtlasRefCode{ANA-TOPQ-2023-35}

\PreprintIdNumber{CERN-EP-2026-268}

\AtlasJournal{JHEP}
\AtlasCoverEgroupAnalysisTeam{}

\hypersetup{pdftitle={ATLAS document},pdfauthor={The ATLAS Collaboration}}

\begin{document}

\maketitle

\tableofcontents

%
%

%
%

% The next lines are included from the .//sections/Introduction.tex input file
\section{Introduction}

In the Standard Model~(SM),
the flavour of both charged and neutral leptons is conserved~\cite{Weinberg:1967tq}. However, the observation of neutrino oscillations~\cite{Maki:1962mu,Pontecorvo:1967fh,Super-Kamiokande:1998kpq,SNO:2001kpb,SNO:2002tuh} implies non-zero neutrino masses and mixing, demonstrating that neutral lepton flavour is not conserved in nature. While the mechanism responsible for neutrino mass generation remains unknown, it is natural to ask whether charged lepton flavour conservation is also violated at an observable level. The simplest extensions of the SM accommodating neutrino masses predict charged lepton flavour violation~(\CLFV), but at a level far below current experimental sensitivity (the decay $\mu \rightarrow e\gamma$ that occurs through neutrino mixing in a $W$ boson loop has a predicted branching fraction of order $10^{-54}$~\cite{Calibbi:2017uvl}).
The observation of \CLFV would therefore provide clear evidence for physics beyond the Standard Model~(BSM), in particular beyond the minimal extensions accommodating neutrino masses.

Models that can generate \CLFV include the unification of leptons and quarks into representations of the $\textrm{SU(5)}$~\cite{SU5}, $\textrm{SO(10)}$~\cite{Georgi:1974my,Fritzsch:1974nn} or $\textrm{SU(5)}\times \textrm{U(1)}$~\cite{SU5U1_1,SU5U1_2,SU5U1_3,SU5U1_4} groups, supersymmetric scenarios~\cite{SUSY1,SUSY2,SUSY3,SUSY4,SUSY5} and technicolor models~\cite{TC1,TC2,TC3}. In particular, some leptoquark models~\cite{LQreview} that accommodate the apparent deviations from SM predictions seen in measurements of $R(D)$ and $R(D^{*})$ \cite{BaBar:2012obs,BaBar:2013mob,Belle:2015qfa,Belle:2016dyj,Belle:2017ilt,Belle:2019rba,LHCb:2023zxo,LHCb:2023uiv,Belle-II:2024ami,HFLAV:2024ctg} in $B$-meson decays can generate \CLFV with a rate as high as $\mathcal{B}(t \to \ell\ell^{'} c) \approx 10^{-6}$~\cite{Kim:2018oih}, where $\ell,\ell^{'} = \{e, \mu, \tau\}$ and $\ell \neq \ell^{'}$.

Processes involving \CLFV have been investigated by several experiments at both low and high energies~\cite{Davidson:2022jai,Calibbi:2017uvl}. ATLAS has contributed to the experimental effort by searching for \CLFV in the decays of the $\tau$-lepton~\cite{EXOT-2014-14,BPHY-2018-05-custom}, $Z$ boson~\cite{EXOT-2013-02, EXOT-2016-36,EXOT-2018-35,EXOT-2020-28,EXOT-2018-36}, Higgs boson~\cite{HIGG-2014-08,HIGG-2015-09,HIGG-2017-08,HIGG-2018-58}, and exotic processes~\cite{EXOT-2019-20}. More recently, the ATLAS and CMS Collaborations have searched for \CLFV in top quark interactions in a variety of final states. The \CLFV interaction is classified according to the flavours of the contributing leptons, and is dependent on different BSM operators for each flavour pair. An ATLAS analysis~\cite{TOPQ-2023-23-custom} targeted the $\mu\tau$ \CLFV process in conjunction with a semileptonic decay of the SM top, while the CMS analysis of the $\mu\tau$ interaction~\cite{CMS-TOP-22-011} considered the hadronic decay channel for the SM top quark; the latter showed greater sensitivity to the $\mu\tau tq$ interaction through its higher statistical precision.
CMS has investigated the $e\mu$ \CLFV process in two different scenarios, in which the decay of the SM top quark via a $W$ boson proceeds either hadronically~\cite{CMS-TOP-19-006} or leptonically~\cite{CMS-TOP-22-005}; here the leptonic channel shows greater sensitivity due to the distinctive three-lepton signature and lower background compared with the hadronic channel. All these searches for \CLFV in top quark processes show results consistent with the SM expectations, allowing limits to be set on \CLFV operators and on branching fractions for \CLFV top quark decays.

\begin{figure}[tb]
\centering
{\centering
\subfloat[]{\includegraphics[width=0.31\textwidth]{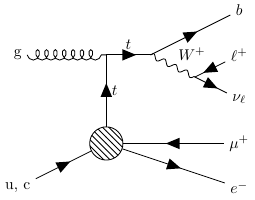}\label{fig:leptonic_production_tchannel_signal_feynman}}\hfill
\subfloat[]{\includegraphics[width=0.31\textwidth]{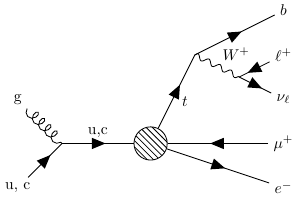}\label{fig:leptonic_production_schannel_signal_feynman}}\hfill
\subfloat[]{\includegraphics[width=0.33\textwidth]{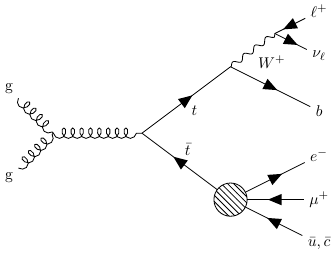}\label{fig:leptonic_decay_signal_feynman}}
}
\caption{Illustrative Feynman diagrams for the process under study, where the hashed circle represents the cLFV vertex: \protect\subref{fig:leptonic_production_tchannel_signal_feynman} single-top-quark production with a cLFV process in the $t$-channel; \protect\subref{fig:leptonic_production_schannel_signal_feynman} single-top-quark production
with a cLFV process in the $s$-channel; and
\protect\subref{fig:leptonic_decay_signal_feynman} top-quark pair production with a cLFV top-quark decay process. The charge conjugate of each process is also considered. Here $\ell \in \{e,\mu,\tau\}$.}
\end{figure}

This analysis targets the $e\mu tq$ \CLFV interaction, by searching for single-top production via the process $gq \rightarrow t\ell^{\pm}\ell'^{\mp}$, where $q = \{u,c\}$ and $\ell\ell' = \{e\mu,\mu e\}$, and the corresponding decay $t \rightarrow \ell^{\pm}\ell'^{\mp}q$ in top-quark pair production ($t\bar{t}$) events, where the other top-quark of the pair decays according to the SM\@. In both cases, the SM-decaying top quark is required to decay into a $b$-quark and a $W$ boson, with the $W$ boson subsequently decaying into an electron or a muon and a neutrino (including indirect decays via a $\tau$-lepton). The production processes are shown in Figures \ref{fig:leptonic_production_tchannel_signal_feynman} and \ref{fig:leptonic_production_schannel_signal_feynman}, and the decay process in Figure \ref{fig:leptonic_decay_signal_feynman}. This search is performed using the full proton--proton~($pp$) collision data sample collected with the ATLAS detector at $\sqrt{s} = 13$~\TeV in 2015--2018 during Run~2 of the Large Hadron Collider~(LHC) \cite{Evans:2008zzb}. This corresponds to an integrated luminosity of 140 fb$^{-1}$.

Provided that the experimentally accessible energy scale is significantly lower than the scale of new physics, the effects of such new physics can be described in a model-independent way within an Effective Field Theory~(EFT) framework.
Here some set of operators $\{\mathcal{O}_i\}$ with dim$(\mathcal{O}_i)>4$ are added to the SM Lagrangian $\mathcal{L}_{\mathrm{SM}}$ according to

\begin{equation}
\mathcal{L} = \mathcal{L}_{\mathrm{SM}} + \mathcal{L}_{\mathrm{eff}} = \mathcal{L}_{\mathrm{SM}} + \sum\limits_{i} \frac{C_{i}}{\Lambda^{2}}\mathcal{O}_i + \cdots,
\end{equation}

with $\Lambda$ the mass scale of the new physics, and Wilson coefficients $C_i$ characterising the strength of each operator. The higher dimensional counter-terms required for a renormalised theory are suppressed by increasing powers of $\Lambda$,
ensuring that the amplitudes at a given order in the expansion remain finite.

The \CLFV production and decay processes may be generated by the set of $\textrm{SU(3)}_{\textrm{C}}\times \textrm{SU(2)}_{\textrm{L}} \times \textrm{U(1)}_{\textrm{Y}}$ dimension-six two-quark two-lepton~(2Q2L) EFT operators listed in~\cref{tab:EFTop}. This list includes all 2Q2L operators that could contribute, and forms a subset of the Warsaw basis operators~\cite{Grzadkowski:2010es}. These operators are combined by Lorentz structure according to~\cref{tab:EFT_combination}. This gives a set of six operators (three for each light quark flavour) and corresponding Wilson coefficients, under the assumptions that the Wilson coefficients are real and invariant with respect to the ordering of the leptons and quarks: $C_{ijkl} = C_{ijkl}^{\dagger} = C_{jilk}$. Here the indices $i,j = \{1,2,3\}$ represent the lepton flavour generations and $k,l = \{1,2,3\}$ represent the quark flavour generations. As \CLFV processes are forbidden in the SM, there is no interference between the SM and EFT amplitudes. Therefore, both the \CLFV production-process cross-section and \CLFV decay partial width $\Gamma(t \to \ell\ell^{'} q)$ grow quadratically with the corresponding Wilson coefficients.  Non-zero Wilson coefficients lead to a large \CLFV production cross-section, whereas the decay branching ratio is significantly suppressed by the size of the top-quark mass relative to the new physics scale, with the decay width at leading order~(LO) given in terms of the Wilson coefficients as~\cite{Chala:2018agk}

\begin{align}\label{eq:master}
\Gamma(t\to\ell^{+}_{i}\ell^{-}_{j}q_{k}) = \frac{m_{t}}{6144\pi^{3}}\left(
\frac{m_{t}}{\Lambda}\right)^{4} \left\{
16|C_{ijk3}^{\mathrm{vector}}|^{2} +
2|C_{ijk3}^{\mathrm{scalar}}|^{2} +
96|C_{ijk3}^{\mathrm{tensor}}|^{2} \right\} \;,
\end{align}

where $m_t$ is the mass of the top quark.

\begin{table}[tb]
\centering
\caption{EFT operator basis. In the convention used, $\mathsf{l}$ and $\mathsf{q}$ are the left-handed lepton and quark doublets, respectively, while $\mathsf{u}$ and $\mathsf{e}$ are the right-handed up-type quark and charged-lepton singlets, respectively. The indices $i,j = \{1,2,3\}$ represent the lepton flavour generations and $k,l = \{1,2,3\}$ represent the quark flavour generations, respectively. The $\sigma^{I}$ are the Pauli matrices, $\varepsilon=i\sigma^{2}$ is the anti-symmetric $SU(2)$ tensor, $\sigma^{\mu\nu}=\frac{i}{2}[\gamma^{\mu},\gamma^{\nu}]$ and $\gamma^{\mu}$ the Dirac matrices. The $^{\ddagger}$ symbol marks non-hermitian operators.}
\label{tab:EFTop}

\begin{tabular}{l|ll}
\toprule
\textbf{Operator} & \textbf{Interaction} & \textbf{Lorentz Structure} \\
\midrule
$\mathcal{O}_{\mathsf{lq}}^{1(ijkl)}$   & $(\bar{\mathsf{l}}_{i}\Gamu \mathsf{l}_{j})(\bar{\mathsf{q}}_{k}\gamu \mathsf{q}_{l})$ & Vector \\[5pt]
$\mathcal{O}_{\mathsf{lq}}^{3(ijkl)}$   & $(\bar{\mathsf{l}}_{i}\Gamu\sigma^{I}\mathsf{l}_{j})(\bar{\mathsf{q}}_{k}\gamu\sigma_{I}\mathsf{q}_{l})$ & Vector \\[5pt]
$\mathcal{O}_{\mathsf{eq}}^{(ijkl)}$    & $(\bar{\mathsf{e}}_{i}\Gamu \mathsf{e}_{j})(\bar{\mathsf{q}}_{k}\gamu \mathsf{q}_{l})$ & Vector \\[5pt]
$\mathcal{O}_{\mathsf{lu}}^{(ijkl)}$    & $(\bar{\mathsf{l}}_{i}\Gamu \mathsf{l}_{j}) (\bar{\mathsf{u}}_{k}\gamu \mathsf{u}_{l})$ & Vector  \\[5pt]
$\mathcal{O}_{\mathsf{eu}}^{(ijkl)}$    & $(\bar{\mathsf{e}}_{i}\Gamu \mathsf{e}_{j})(\bar{\mathsf{u}}_{k}\gamu \mathsf{u}_{l})$ & Vector  \\[5pt]
$^{\ddagger}\mathcal{O}_{\mathsf{lequ}}^{1(ijkl)}$ & $(\bar{\mathsf{l}}_{i}\mathsf{e}_{j})\varepsilon(\bar{\mathsf{q}}_{k} \mathsf{u}_{l})$ & Scalar \\[5pt]
$^{\ddagger}\mathcal{O}_{\mathsf{lequ}}^{3(ijkl)}$ & $(\bar{\mathsf{l}}_{i}\sigma^{\mu\nu}\mathsf{e}_{j})\varepsilon(\bar{\mathsf{q}}_{k}\sigma_{\mu\nu} \mathsf{u}_{l})$ & Tensor \\[5pt]
\bottomrule
\end{tabular}
\end{table}

\begin{table}[tb]
\centering
\caption{Combinations of the \CLFV generating operators outlined in \cref{tab:EFTop}, sorted by light quark flavour and Lorentz structure, which are individually treated in this analysis. The symbols $\sum ijkl$ refer to a sum over the permutations $ijkl ~ + ~ jikl ~ + ~ ijlk ~ + ~ jilk$, with $i,j \in \{1,2\}$ corresponding to $\ell \in \{e,\mu\}$ and $k,l \in \{1,2\}$ corresponding to $q \in \{u,c\}$. The operator $\mathcal{O}_{lq}^{-(ijkl)}$ is the linear combination $\mathcal{O}^{-(ijkl)}_{\mathsf{lq}} = \mathcal{O}_{\mathsf{lq}}^{1(ijkl)} - \mathcal{O}_{\mathsf{lq}}^{3(ijkl)}$, which contains the interactions of two up-type quarks with two charged leptons, the orthogonal combination $\mathcal{O}^{(ijkl)}_{\mathsf{lq}} = \mathcal{O}_{\mathsf{lq}}^{1(ijkl)} + \mathcal{O}_{\mathsf{lq}}^{3(ijkl)}$ contains the interactions of two up-type quarks and two neutrinos and so cannot generate \CLFV at tree level.}
\begin{tabular}{c|c|c|l}
\toprule
\makecell{\textbf{Light quark} \\ \textbf{flavour}} & \makecell{\textbf{Lorentz} \\ \textbf{structure}}  & \makecell{\textbf{Wilson} \\ \textbf{coefficient}} & \textbf{Operator} \\
\midrule
$u$ & Tensor  & $C^{\mathrm{tensor}}_{e\mu tu}$ & $\mathcal{O}^{\mathrm{tensor}}_{e\mu tu} = \mathcal{O}^{3(\sum 1213)}_{\mathsf{lequ}} + {\mathcal{O}^{3(\sum 1213)}_{\mathsf{lequ}}}^{\dagger}$\\[5pt]
& Vector  & $C_{e \mu tu}^{\mathrm{vector}}$ & $\mathcal{O}_{e \mu t u}^{\mathrm{vector}} = \mathcal{O}^{-(\sum 1213)}_{\mathsf{lq}} + \mathcal{O}^{(\sum 1213)}_{\mathsf{eq}} + \mathcal{O}^{(\sum 1213)}_{\mathsf{lu}} + \mathcal{O}^{(\sum 1213)}_{\mathsf{eu}}$ \\[5pt]
& Scalar  & $C_{e \mu tu}^{\mathrm{scalar}}$ & $\mathcal{O}_{e \mu t u}^{\mathrm{scalar}} = \mathcal{O}^{1(\sum 1213)}_{\mathsf{lequ}} + {\mathcal{O}^{1(\sum 1213)}_{\mathsf{lequ}}}^{\dagger}$\\
\midrule
$c$ & Tensor  &  $C^{\mathrm{tensor}}_{e\mu tc}$   &  $\mathcal{O}^{\mathrm{tensor}}_{e\mu tc} = \mathcal{O}^{3(\sum 1223)}_{\mathsf{lequ}} + {\mathcal{O}^{3(\sum 1223)}_{\mathsf{lequ}}}^{\dagger}$ \\[5pt]
& Vector  &   $C^{\mathrm{vector}}_{e \mu tc}$ & $\mathcal{O}_{e \mu tc}^{\mathrm{vector}} = \mathcal{O}^{-(\sum 1223)}_{\mathsf{lq}} + \mathcal{O}^{(\sum 1223)}_{\mathsf{eq}} + \mathcal{O}^{(\sum 1223)}_{\mathsf{lu}} + \mathcal{O}^{(\sum 1223)}_{\mathsf{eu}}$  \\[5pt]
& Scalar  &    $C^{\mathrm{scalar}}_{e \mu tc}$  & $\mathcal{O}_{e \mu tc}^{\mathrm{scalar}} = \mathcal{O}^{1(\sum 1223)}_{\mathsf{lequ}} + {\mathcal{O}^{1(\sum 1223)}_{\mathsf{lequ}}}^{\dagger}$ \\
\bottomrule
\end{tabular}
\label{tab:EFT_combination}
\end{table}

In this analysis, the measured data are interpreted within this framework, allowing determination of the moduli of the Wilson coefficients corresponding to each operator in \cref{tab:EFT_combination}, or for upper limits to be placed in the absence of any significant deviation from the SM expectation.

% End of text imported from the .//sections/Introduction.tex input file

%

%

% The next lines are included from the .//sections/ATLASDetector.tex input file
\section{ATLAS detector}

\newcommand{\AtlasCoordFootnote}{%
ATLAS uses a right-handed coordinate system with its origin at the nominal interaction point (IP)
in the centre of the detector and the \(z\)-axis along the beam pipe.
The \(x\)-axis points from the IP to the centre of the LHC ring,
and the \(y\)-axis points upwards.
Polar coordinates \((r,\phi)\) are used in the transverse plane,
\(\phi\) being the azimuthal angle around the \(z\)-axis.
The pseudorapidity is defined in terms of the polar angle \(\theta\) as \(\eta = -\ln \tan(\theta/2)\) and is equal to the rapidity
$ y = \frac{1}{2} \ln \left( \frac{E + p_z}{E - p_z} \right) $ in the relativistic limit.
Angular distance is measured in units of \(\Delta R \equiv \sqrt{(\Delta y)^{2} + (\Delta\phi)^{2}}\).}

The ATLAS detector~\cite{PERF-2007-01} at the LHC covers nearly the entire solid angle around the collision point.\footnote{\AtlasCoordFootnote}
It consists of an inner tracking detector surrounded by a thin superconducting solenoid, electromagnetic and hadronic calorimeters,
and a muon spectrometer incorporating three large superconducting air-core toroidal magnets.

The inner-detector system (ID) is immersed in a \qty{2}{\tesla} axial magnetic field
and provides charged-particle tracking in the range \(|\eta| < 2.5\).
The high-granularity silicon pixel detector covers the vertex region and typically provides four measurements per track,
the first hit generally being in the insertable B-layer (IBL) installed before Run~2~\cite{ATLAS-TDR-19,PIX-2018-001}.
It is followed by the SemiConductor Tracker (SCT), which usually provides eight measurements per track.
These silicon detectors are complemented by the transition radiation tracker (TRT),
which enables radially extended track reconstruction up to \(|\eta| = 2.0\).
The TRT also provides electron identification information
based on the fraction of hits (typically 30 in total) above a higher energy-deposit threshold corresponding to transition radiation.

The calorimeter system covers the pseudorapidity range \(|\eta| < 4.9\).
Within the region \(|\eta|< 3.2\), electromagnetic calorimetry (ECAL) is provided by barrel and
endcap high-granularity lead/liquid-argon (LAr) calorimeters,
with an additional thin LAr presampler covering \(|\eta| < 1.8\)
to correct for energy loss in material upstream of the calorimeters.
Hadronic calorimetry is provided by the steel/scintillator-tile calorimeter,
segmented into three barrel structures within \(|\eta| < 1.7\), and two copper/LAr hadronic endcap calorimeters.
The solid angle coverage is completed with forward copper/LAr and tungsten/LAr calorimeter modules
optimised for electromagnetic and hadronic energy measurements, respectively.

The muon spectrometer (MS) comprises separate trigger and
high-precision tracking chambers measuring the deflection of muons in a magnetic field generated by the superconducting air-core toroidal magnets.
The field integral of the toroids ranges between \num{2.0} and \qty{6.0}{\tesla\metre}
across most of the detector.
Three layers of precision chambers, each consisting of layers of monitored drift tubes, cover the region \(|\eta| < 2.7\),
complemented by cathode-strip chambers in the forward region, where the background is highest.
The muon trigger system covers the range \(|\eta| < 2.4\) with resistive-plate chambers in the barrel, and thin-gap chambers in the endcap regions.

The luminosity is measured mainly by the LUCID--2~\cite{LUCID2} detector that records Cherenkov light produced in the quartz windows of photomultipliers located close to the beampipe.

Events were selected by the first-level trigger system implemented in custom hardware,
followed by selections made by algorithms implemented in software in the high-level trigger~\cite{TRIG-2016-01}.
The first-level trigger accepted events from the \qty{40}{\MHz} bunch crossings at a rate close to \qty{100}{\kHz},
which the high-level trigger further reduced in order to record complete events to disk at about \qty{1.25}{\kHz}.

A software suite~\cite{SOFT-2022-02} is used in data simulation, in the reconstruction
and analysis of real and simulated data, in detector operations, and in the trigger and data acquisition
systems of the experiment.

% End of text imported from the .//sections/ATLASDetector.tex input file

%

%

% The next lines are included from the .//sections/AnalysisStrategy.tex input file
\section{Analysis strategy}

An overview of the analysis strategy is given here, before each step is described in more detail in the following sections.

Selected events are separated into five categories, called analysis regions, based on the flavours of the charged leptons, the quality of their reconstruction and their isolation from other activity in the event. Two of these regions are \SRs\ (SRs). The remaining three are high-purity \CRs\ (CRs) that target the main backgrounds that can enter the SRs. These backgrounds either contain only prompt leptons, which are produced in the initial interaction through the decay of electroweak bosons, or contain non-prompt or fake leptons as well.
The non-prompt leptons are produced in the decays of secondary particles, while fake leptons originate from other particles misidentified in the detector. One CR targets the prompt-lepton $WZ$ background. The other two CRs are intended to control the background contribution from
non-prompt leptons originating from heavy-flavour (HF) hadron decays inside jets.

The two SRs are designed to select the \CLFV signatures with lepton flavours $(e\mu)e$ or $(e\mu)\mu$, where the $(e\mu)$ pair originates from the \CLFV vertex and the remaining charged lepton is produced in a SM top-quark decay.
One of the SRs (SR-decay) targets \CLFV in top-quark decays in \ttbar events, and the other (SR-prod) targets \CLFV in single top-quark production.
Kinematic variables are then used as inputs to multivariate discriminators (Boosted Decision Trees), which are separately trained and deployed in both SRs, to discriminate between signal and background events.  Further details are given in~\cref{sec:BDT}.

The signal contribution in the SRs is then determined with a binned profile-likelihood fit, as described in~\cref{sec:results}.  The fit is performed on the Boosted Decision Tree (BDT) output distribution in the SR-decay and SR-prod regions, with the CRs included in the fit to simultaneously determine the normalisations of their targeted backgrounds. Systematic uncertainties are considered through nuisance parameters in the fits.

An EFT interpretation is performed by fitting to samples generated with a single active EFT operator; the fit is repeated for each operator in turn. The values of the Wilson coefficients extracted from each fit can be converted into branching ratios using \cref{eq:master}.

% End of text imported from the .//sections/AnalysisStrategy.tex input file

%

%

% The next lines are included from the .//sections/DataMC.tex input file
\section{Data and simulated event samples}

The results in this paper use $pp$ collision data recorded with the ATLAS detector during Run 2 at the LHC (2015--2018), at a centre-of-mass energy of $\sqrt{s} = 13$~\TeV.
After requiring all detector subsystems to be operational under stable beam conditions~\cite{DAPR-2018-01}, the data sample corresponds to an integrated luminosity of 140~fb$^{-1}$, as determined using the LUCID-2 detector~\cite{LUCID2,DAPR-2021-01} for the primary luminosity measurements.

Monte Carlo (MC) simulated event samples are used to model the signal and background processes. The effect of multiple interactions in the same and neighbouring bunch
crossings (\pileup) was modelled by overlaying the simulated hard-scattering event with
inelastic \(pp\) events generated with \PYTHIA[8.186]~\cite{Sjostrand:2007gs}
using the \NNPDF[2.3lo] set of parton distribution functions (PDF)~\cite{Ball:2012cx} and the
A3 set of tuned parameters~\cite{ATL-PHYS-PUB-2016-017}. Following the event generation, the ATLAS detector response was simulated~\cite{SOFT-2010-01} with the \GEANT toolkit~\cite{Agostinelli:2002hh}.  Some samples used in the estimation of systematic uncertainties were produced using a fast-simulation package that relies on a parameterisation of the calorimeter response~\cite{SOFT-2010-01}.
All samples of simulated events, except those produced with the \SHERPA~\cite{Bothmann:2019yzt} generator, use \EVTGEN~\cite{Lange:2001uf}
to model the decays of bottom and charm hadrons.
The top-quark mass was set to $172.5~\GeV$.

The EFT signal processes are simulated as $gq_{k} \to t\ell_i\ell_j$ and $\ttbar \to (\ell_i\ell_jq_{k})(W(\to \ell\nu) b)$ for the production and decay diagrams, respectively, with $\ell_i, \ell_j = \{e, \mu, \tau\}$ and $q_{k} = \{u,c\}$; however, only $\ell_i,\ell_j \in \{e,\mu\}$ are considered in this analysis. A UFO model~\cite{ufo} containing the EFT operators listed in \cref{tab:EFTop} was created with \FEYNRULES[2.0]~\cite{feynrules}, using the \textsc{dim6top} model~\cite{AguilarSaavedra:2018nen}. Events are generated at LO in Quantum Chromodynamics (QCD) with \MADGRAPH{2.9.5}~\cite{Alwall:2014hca} for the hard process, in conjunction with \PYTHIA[8.306]~\cite{Bierlich:2022pfr} for showering and hadronisation.
The dynamic renormalisation and factorisation scales (\muR, \muF) correspond to the centre-of-mass energy of the incoming partons for the top-quark decay process, and half the sum of the transverse masses of all final state particles and partons for the single-top production process.
Events are generated in the five-flavour scheme, in which all the quark masses are set to zero, except for the top quark mass.  The \NNPDF[3.1nlo]~\cite{NNPDF:2017mvq} PDF is used, and \PYTHIA{8} is configured according to the A14 tune~\cite{ATL-PHYS-PUB-2014-021}. This signal model is the same as that used in the ATLAS analysis~\cite{TOPQ-2023-23-custom} targeting the $\mu \tau tq$ \CLFV process, apart from the lepton flavours involved, and is consistent with the LO order model used in the CMS analysis~\cite{CMS-TOP-22-005} targeting the same $e\mu tq$ process considered here.

Cross-sections for the single-top \CLFV production processes are determined directly from the generator.
Those for the \CLFV \ttbar decay samples are evaluated according to

\begin{equation}
\sigma^{\textrm{cLFV}}_{ijk} = 2\cdot\sigma_{\ttbar}\cdot\mathcal{B}(t\to W(\to\ell\nu)b)\cdot\frac{\Gamma(t\to\ell^{+}_{i}\ell^{-}_{j}q_{k})}{\Gamma(t)}\, .
\end{equation}

Here, $\sigma_{\ttbar}$ is the \ttbar production cross-section prediction at next-to-next-to-leading order (NNLO) in QCD, including the resummation of next-to-next-to-leading logarithmic (NNLL) soft-gluon terms calculated using \TOPpp[2.0]~\cite{Czakon:2011xx}, and is equal to $833.9$ pb.\footnote{The fractional scale and PDF uncertainties evaluated from the LO signal generator are propagated to the nominal NNLO+NNLL normalisation and are used throughout the analysis, including in the fits.}

Cross-sections corresponding to the tensor, vector, and scalar Wilson coefficients that contribute to the \CLFV $\ell\ell^{'}tq$ process are given in~\cref{tab:clfv_xsecs_all}, summed over all lepton generations ($\ell,\ell^{'} \in \{e,\mu,\tau\}$ where $\ell \neq \ell^{'}$). These assume a top-quark mass of $172.5$~\GeV, a top-quark decay width $\Gamma(t)$ of $1.32$~\GeV at NNLO and a branching ratio ${\cal  B}(t\to W(\ell\nu)b) = 0.324$. The scale of new physics is set to $\Lambda=1$~\TeV and a value of 1.0 is assumed for the relevant Wilson coefficient, with others set to zero.
The production cross-sections are quoted for up- and charm-quark couplings separately ($k = 1,2$) and differ due to the proton PDFs\@.
The \ttbar cross-sections involving $\ell\ell^{'}q_{k}$ decays are quoted for $q_k = u,c$ together, as the difference between the two light-quark flavours is negligible for the decay process considered.

\begin{table}[tb]
\centering
\caption{Theoretical cross-sections for single top-quark production and top-quark decays through \CLFV $\ell \ell^{'} t q_k$ interactions involving vector, scalar, and tensor EFT operators, in $\sqrt{s} = 13$~\TeV $pp$ collisions. Here $\ell,\ell^{'} \in \{e,\mu,\tau\}$, $\ell \neq \ell^{'}$, and the cross-sections are summed over all lepton generations. The single top-quark production cross-sections are given for $u$- and $c$-quark couplings separately, but are combined for the $\ttbar$ decay process ($q_k = u,c$). For the calculation of the cross-section the value of the relevant Wilson coefficient is set to 1, and the scale of new physics is set to $\Lambda = 1$~\TeV. All cross-sections include the leptonic decay of the SM-decaying top quark, assuming a branching ratio $\mathcal{B}(t\rightarrow W(\ell \nu)b) = 0.324$. Scale and PDF uncertainties are shown for each cross-section.}
\setlength\extrarowheight{5pt}
\small
\begin{tabular}{c|cccccccc}
\toprule
& \multicolumn{3}{c}{\textbf{Cross-section $\sigma \pm \delta\sigma(\textrm{Scale}) \pm \delta\sigma(\textrm{PDF})$ (fb)}} \\
& Vector & Scalar & Tensor \\ \midrule
\textbf{Production $\ell\ell^{'}tu$} & $472^{+95}_{-76}\pm 5.6$ & $101^{+21}_{-16}\pm 1$ & $2150^{+410}_{-320}\pm 20$ \\
\textbf{Production $\ell\ell^{'}tc$} & $32^{+4.9}_{-4.1}\pm 6.5$ & $6.1^{+1.0}_{-0.8}\pm 1.5$ & $153^{+21}_{-18}\pm 29$ \\
\textbf{Decay $\ell\ell^{'}tq_{k}$} & $32^{+8.2}_{-6.0}\pm 0.3$ & $3.9^{+1.0}_{-0.7}\pm 0.03$ & $189^{+49}_{-36}\pm 2$ \\
\bottomrule
\end{tabular}
\label{tab:clfv_xsecs_all}
\end{table}

Several sources of background are considered, including those with a genuine signature of three or more prompt, charged leptons (e.g. diboson, \ttW, \ttZ or \ttH), and those with two prompt leptons and a fake or non-prompt lepton (e.g. \ttbar, single top and \Zjets).  Unless otherwise specified, these MC samples use \PYTHIA[8] to model the parton shower (PS),
hadronisation, and underlying event, with parameters set according
to the A14 tune
and using the \NNPDF[2.3lo]
set of PDFs~\cite{Ball:2012cx}.

The production of \ttbar events was modelled using the
\POWHEGBOX[v2]~\cite{Frixione:2007nw,Nason:2004rx,Frixione:2007vw,Alioli:2010xd}
generator at next-to-leading order (NLO) with the \NNPDF[3.0nlo]~\cite{Ball:2014uwa} PDF set
and the \hdamp parameter set
to 1.5\,\mtop~\cite{ATL-PHYS-PUB-2016-020}.\footnote{The
\hdamp parameter is a resummation damping factor and one of the
parameters that controls the matching of \POWHEG matrix elements to
the parton shower and thus effectively regulates the
high-\pT radiation against which the \ttbar system recoils.}   The events were interfaced
to \PYTHIA[8.230]~\cite{Sjostrand:2014zea}.  The \ttbar sample was normalised to the same NNLO+NNLL cross-section as the \CLFV signal decay sample.

The associated production of top quarks with \(W\) bosons (\(tW\)) was
modelled by the
\POWHEGBOX[v2]~\cite{Re:2010bp}
generator at NLO in QCD using the five-flavour scheme and the
\NNPDF[3.0nlo] set of PDFs~\cite{Ball:2014uwa}.
The diagram removal scheme~\cite{Frixione:2008yi} was used to
remove interference and overlap with \ttbar production.
The related uncertainty was estimated by comparison with an alternative sample
generated using the diagram subtraction scheme~\cite{Frixione:2008yi,ATL-PHYS-PUB-2016-020}.
The events were interfaced to \PYTHIA[8.230].
The predicted cross-section was computed at NLO in QCD with the addition of third-order corrections to soft-gluon emissions by resumming NNLL terms~\cite{Kidonakis:2021vob}.

The production of single-top \tZq\ and \tWZ\ events was modelled using the NLO generator \MGNLO[2.3.3]~\cite{Alwall:2014hca}
with the \NNPDF[3.0nlo] PDF, interfaced to \PYTHIA[8.230] for \tZq\ and to \PYTHIA[8.212] for \tWZ; their cross-sections are taken from the generator.

The associated production of a $\ttbar$ pair with a leptonically-decaying \Zboson boson (\ttZ) was modelled using the \MGNLO[2.8.1] generator at NLO with the
\NNPDF[3.0nlo] PDF, interfaced to \PYTHIA[8.244], while the production of \ttH events was modelled using the
\POWHEGBOX[v2]
generator at NLO with the \NNPDF[3.0nlo] PDF set, interfaced to \PYTHIA[8.230].
The cross-sections for \ttZ and \ttH were calculated at NLO QCD and NLO electroweak (EW) accuracy using
\MGNLO as reported in Ref.~\cite{deFlorian:2016spz}.

The production of $\ttbar W$ events was simulated at NLO precision in QCD with \SHERPA[2.2.10] ~\cite{Bothmann:2019yzt}
and the \NNPDF[3.0nnlo] PDF set~\cite{Ball:2014uwa}.
In this set-up, matrix elements (MEs) with different parton multiplicities were matched and merged with the \SHERPA PS model based on the Catani--Seymour dipole factorisation scheme~\cite{Gleisberg:2008fv,Schumann:2007mg}.
The virtual QCD corrections (in the coupling $\alphas$) for MEs at NLO accuracy were provided by the \OPENLOOPS library~\cite{Buccioni:2019sur,Cascioli:2011va,Denner:2016kdg}. Up to one additional parton was included in
the NLO ME, and up to two additional partons were included at LO in QCD using Comix~\cite{Gleisberg:2008fv}. The merging scale parameter~(\muQ), which sets a threshold to determine what part of the phase-space
is filled by the PS or the ME generator, was set to an energy of \SI{30}{\GeV}.
Following the same procedure as in Ref.~\cite{TOPQ-2023-23-custom},
sub-leading EW corrections were
partially accounted for via the addition of an independent \Sherpa[2.2.10] sample produced at LO in QCD for this final state.
The cross-section used for $t\bar{t}W$ QCD+EW production is $745\,\mathrm{fb}$, calculated at NNLO QCD and including NLO EW corrections~\cite{Buonocore:2023ljm}.

Samples of diboson final states (\(VV\)) were simulated with the \SHERPA[2.2.2] generator
including off-shell effects and Higgs boson contributions, where appropriate.
Fully leptonic final states and semileptonic final states, where one boson
decays leptonically and the other hadronically, were generated using
MEs at NLO accuracy in QCD for up to one additional parton
and at LO accuracy for up to three additional parton
emissions. Samples for the loop-induced processes \(gg \to VV\) were
generated using LO-accurate MEs for up to one
additional parton emission.
The ME calculations were matched
and merged with the \SHERPA PS, and the
\NNPDF[3.0nnlo] set of PDFs was used.

The production of \(Z+\)jets was simulated with the
\SHERPA[2.2.11] or 2.2.14
generator using NLO MEs for up to two partons, and LO MEs
for up to five partons calculated with the Comix
and \OPENLOOPS libraries. They
were matched with the \SHERPA PS
and the \NNPDF[3.0nnlo] set of PDFs was used.

The production of triboson (\(VVV\)) events was simulated with the
\SHERPA[2.2.2] generator using factorised gauge-boson decays.
MEs, accurate to NLO for the inclusive process and to LO for up to
two additional parton emissions, were matched and merged with the \SHERPA PS.
Samples were generated using the \NNPDF[3.0nnlo] PDF set.

Other background processes make only minor contributions and are also modelled by MC: $t\bar{t}t\bar{t}$ is simulated with \MGNLO; $t\bar{t}t$ and $t\bar{t}WW$ with \MADGRAPH[2.2.2]; $WH / ZH$ with \POWHEGBOX; \ttgamma with \MADGRAPH[2.3.3] and $Z\gamma$ with \SHERPA[2.2.8]. All samples are interfaced to \PYTHIA[8] except for the $Z\gamma$ sample.
An overlap removal procedure is applied to remove events in the \ttbar and \(Z+\)jets samples that have a photon in the matrix element, to prevent double-counting with the \ttgamma~and $Z\gamma$ samples.

% End of text imported from the .//sections/DataMC.tex input file

%

%
%
%

%

% The next lines are included from the .//sections/ObjectEventSelection.tex input file
\section{Event reconstruction}

Events are required to have a
reconstructed primary vertex, defined as the vertex for which the associated tracks (with $\pT > 500$~\MeV) have the highest sum of $\pT^2$~\cite{IDTR-2021-01}.
Events are selected with  single-electron and single-muon triggers~\cite{TRIG-2016-01,TRIG-2018-05,TRIG-2018-01}.  The lowest-threshold triggers include isolation requirements; these
had \pT thresholds of  20 (24)~\GeV for muons (electrons) in 2015, and thresholds of 26~\GeV in 2016--2018.  They are combined with additional higher-threshold triggers with looser isolation requirements. At least one reconstructed lepton candidate must correspond to a lepton selected by the trigger, with a lepton \pT at least 1~\GeV above the trigger \pT threshold.

Muon candidates are reconstructed by combining ID tracks~\cite{PERF-2015-08} with track segments or full tracks in the MS~\cite{PERF-2015-10,MUON-2022-01,MUON-2018-03}.
Candidates are required to satisfy a \emph{Medium}~\cite{MUON-2018-03} identification criterion and must satisfy $\pt > \SI{10}{\GeV}$ and $|\eta| < 2.5$.
Two muon categories are defined, ``Tight'' and ``Loose'', targeting prompt and non-prompt muons, respectively.
Isolation criteria are applied to the muons, based on a multivariate BDT discriminant, referred to as the non-prompt lepton BDT~\cite{HIGG-2017-02,MUON-2018-03}.  This discriminant combines calorimeter and tracking information to suppress backgrounds from decays of heavy-flavour hadrons and misidentified jets, and has tight and loose requirements.
The ``Tight'' muon candidates must meet the tight isolation criterion, with a transverse impact parameter significance satisfying $|d_0|/\sigma_{d_0} < 3$. ``Loose'' muon candidates are only required to meet the loose isolation criterion, and have an inverted requirement on the transverse impact parameter of $|d_0|/\sigma_{d_0} \geq 3$.

Electron candidates are reconstructed from energy clusters in the ECAL that are associated with charged-particle tracks reconstructed in the ID~\cite{EGAM-2018-01}. Candidates are selected with $\pT > \SI{10}{\GeV}$ and a cluster pseudorapidity satisfying $|\eta| < 2.47$, excluding the transition region between different ECAL components, $1.37 < |\eta| <1.52$. A multivariate likelihood discriminant combining shower-shape and track information is used to distinguish real electrons from hadronic showers; electron candidates are required to meet the \emph{LooseAndBLayerLH} identification criteria~\cite{EGAM-2018-01,PERF-2017-01}.
Two electron categories are then defined, ``Tight'' and ``Loose'', targeting prompt and non-prompt electrons, respectively. The ``Tight'' electron candidates must also meet the \emph{TightLH} identification, the tight isolation criterion from the non-prompt lepton BDT, and a transverse impact parameter significance of $|d_0|/\sigma_{d_0} < 5$. ``Loose'' electron candidates have no additional identification or isolation requirements, and have an inverted requirement on the transverse impact parameter of $|d_0|/\sigma_{d_0} \geq 5$.

The longitudinal impact parameter for either electrons or muons must satisfy $| z_{0} \sin\theta | < \SI{0.5}{mm}$.

Jets are reconstructed using the \antikt algorithm~\cite{Cacciari:2008gp} as implemented in FastJet~\cite{Fastjet} with a jet radius parameter $R=0.4$. The inputs to this algorithm are particle flow (PFlow) objects~\cite{PERF-2015-09}, which combine measurements from the ATLAS inner detector and calorimeters~\cite{PERF-2014-07} to improve the jet energy resolution and increase the jet reconstruction efficiency, especially at low jet \pT. Calibrations are applied to the Jet Energy Scale (JES) and Jet Energy Resolution (JER), which include components derived both from simulation and in situ measurements, as documented in Ref.~\cite{JETM-2018-05}.
Jet candidates are required to have  $\pT> 25$~\GeV and $|\eta|<2.5$.  To reduce the effect of pile-up and to identify the jets that originate from the selected primary vertex, any jet with $\pT<60$~\GeV and $|\eta|<2.4$ must satisfy the \emph{tight} working point of the Jet Vertex Tagger (JVT)~\cite{PERF-2014-03}.

Jets containing $b$-hadrons are identified using the `DL1r' $b$-tagging algorithm~\cite{FTAG-2019-07}. This neural network algorithm combines information from the impact parameters of displaced vertices, in addition to topological properties of secondary and tertiary vertices within a jet. Three scores are determined, corresponding to the probabilities that the given jet corresponds to a light-flavour jet (initiated by $u$-, $d$- or $s$-quarks or gluons), $c$-jet or $b$-jet; the scores are combined into a single discriminant. The analysis uses a 77\% efficiency working point, calibrated using $\ttbar$ events at $\sqrt{s}=13$~\TeV, with rejection factors of approximately 5 and 170 for $c$-jets and light-flavour jets, respectively~\cite{FTAG-2018-01,FTAG-2019-07}.%

The missing transverse momentum in the event, ${\vec{p}_{\textrm{T}}}^\textrm{\,miss}$, whose magnitude is indicated by \MET, is calculated as the negative vector sum of the transverse momentum of all identified hard physics objects (electrons, muons and jets), with a contribution from an additional soft term. This soft term is calculated from ID tracks matched to the primary vertex and not assigned to any of the hard objects~\cite{JETM-2020-03}.

To avoid double-counting detector signatures, reconstructed objects are removed in the following order:
electrons sharing a track with any other electron;
electrons sharing a track with a muon;
jets within $\Delta R = 0.2$ of an electron;
electrons within $\Delta R = 0.4$ of a jet;
jets within $\Delta R = 0.2$ of a muon if they have at most two associated tracks;
muons within $\Delta R = 0.4$ of a jet.

Scale factors (SFs) are used to correct the efficiencies in simulation to match the efficiencies measured in data for the muon~\cite{PERF-2015-10,MUON-2018-03,MUON-2022-01}  and electron~\cite{EGAM-2018-01,PERF-2017-01,EGAM-2021-02} trigger, reconstruction, identification, and isolation criteria.
Additional SFs are applied
for the $b$-tagging efficiencies for jets that originate from the hadronisation of $b$-quarks~\cite{FTAG-2018-01}, $c$-quarks~\cite{FTAG-2020-08}, and light quarks or gluons~\cite{FTAG-2019-02}, and for the JVT pile-up requirement~\cite{PERF-2014-03}.

\section{Event selection}

Selected events are separated into five analysis regions: two SRs targeting the \CLFV signal processes, and three CRs used to constrain the normalisation of prompt and non-prompt lepton backgrounds. All regions used in the analysis are subject to the same
preselection requirements shown in Table \ref{tab:preselection}: at least three Loose leptons, of which at most three are Tight leptons, and at least one jet.
The sum of lepton charges must be $\pm$1, and low mass resonances are vetoed by excluding any opposite-sign same-flavour (OSSF) lepton pair with an invariant mass below 15~\GeV.

\begin{table}[tb]
\centering
\caption{Common preselection criteria for all analysis regions. Leptons are labelled 1, 2 or 3, ordered by decreasing \pT values.}
\begin{tabular}{l|c}
\toprule
\textbf{Description} & \textbf{Preselection}\\
\midrule
Lepton flavours & $ e $ or $\mu$ \\
Baseline lepton definition & $\pt > 10$~\GeV, \ $|\eta|<2.5$ \\
Baseline jet definition & $\pt > 25$~\GeV, \ $|\eta|<2.5$ \\
Number of leptons & $N_{\text{lep}} \geq 3$ \\
Number of Tight leptons & $N_{\text{lep}}^{\text{tight}} \leq 3$ \\
Number of jets & $N_{\mathrm{jets}} \geq 1$ \\
Number of $b$-tagged jets & $N_{b-\mathrm{tags}} \leq 2$ \\
Leading lepton \pt & $\pt^{\ell_1} > 27$~\GeV \\
Sum of lepton charges & $\sum q_{i} = \pm 1$ \\
OSSF lepton pair mass & $m^{\text{OSSF}}_{\ell\ell} \geq 15$~\GeV\\
\bottomrule
\end{tabular}
\label{tab:preselection}
\end{table}

\subsection{Signal Regions}

The SRs require exactly three Tight leptons, with at least one opposite-sign different-flavour (OSDF) $e\mu$ pair, to match the final state lepton topology of the \CLFV signal process. A requirement on the third-leading lepton \pT is used to reduce the non-prompt lepton contribution. To suppress diboson and \Zjets backgrounds, events containing an OSSF dilepton pair within 30~\GeV of the $Z$ boson mass (91.2~\GeV) are removed.

Events are split between two regions, SR-decay and SR-prod, targeting the $t\bar{t}$ decay and single-top production signal channels, respectively. As decay channel events generally contain multiple jets, events with a single jet are placed into SR-prod, with a requirement that \HT (the scalar \pT sum of all final state jets and leptons) is greater than 400~\GeV to suppress the diboson background.

Events with multiple jets are split between SR-decay and SR-prod using a kinematic reconstruction of the ``SM top'' and the lepton-flavour-violating (LFV) top quark, labelled the ``LFV top''.
The SM top candidate is the top quark that decays by the SM leptonic process $t \rightarrow W(\rightarrow \ell\nu)b$, and is present in both decay and production channels. It is reconstructed by assigning the ${\vec{p}_{\textrm{T}}}^\textrm{\,miss}$ to the transverse components of the neutrino four-momentum, and forming $W$ boson candidates from the neutrino and each charged lepton for which the remaining two leptons form an OSDF pair, assuming that the $W$ is on-shell. Due to finite detector resolution, solutions to the transverse $W$ mass $m_{\mathrm{T}}(W)$ can be larger than $m_W$~\footnote{$m_{\mathrm{T}}(W) = \sqrt{2p_{\text{T}}^{\ell}\MET(1-\cos{\Delta\phi})}$, where $\Delta\phi$ is the angle between the lepton and ${\vec{p}_{\textrm{T}}}^\textrm{\,miss}$ in the transverse plane and $p_{\text{T}}^{\ell}$ is the transverse momentum of the lepton.}; in this case, the reconstruction is performed by setting $m_{\mathrm{T}}(W) = m_W$. Top quark candidates are formed from the $W$ candidates and the jet satisfying the tightest \texttt{DL1r} $b$-tagging working point.
If multiple jets satisfy the same working point, the candidate with a mass closest to $m_t$ is chosen as the SM top.
The LFV top is the top quark decaying via the \CLFV process,
and is present only in decay-channel events. It is reconstructed after the SM top, with LFV top candidates formed from the remaining OSDF lepton pair and each remaining jet. The candidate whose mass $m^{\mathrm{LFV}}_t$ is closest to $m_t$ is chosen as the LFV top. The reconstruction is found to have good performance in simulation, with the leptons correctly assigned in $\sim$~90\% of production signal events, and $\sim$~80\% of decay signal events.

\begin{figure}[tb]
\centering
\includegraphics[width=0.5\linewidth]{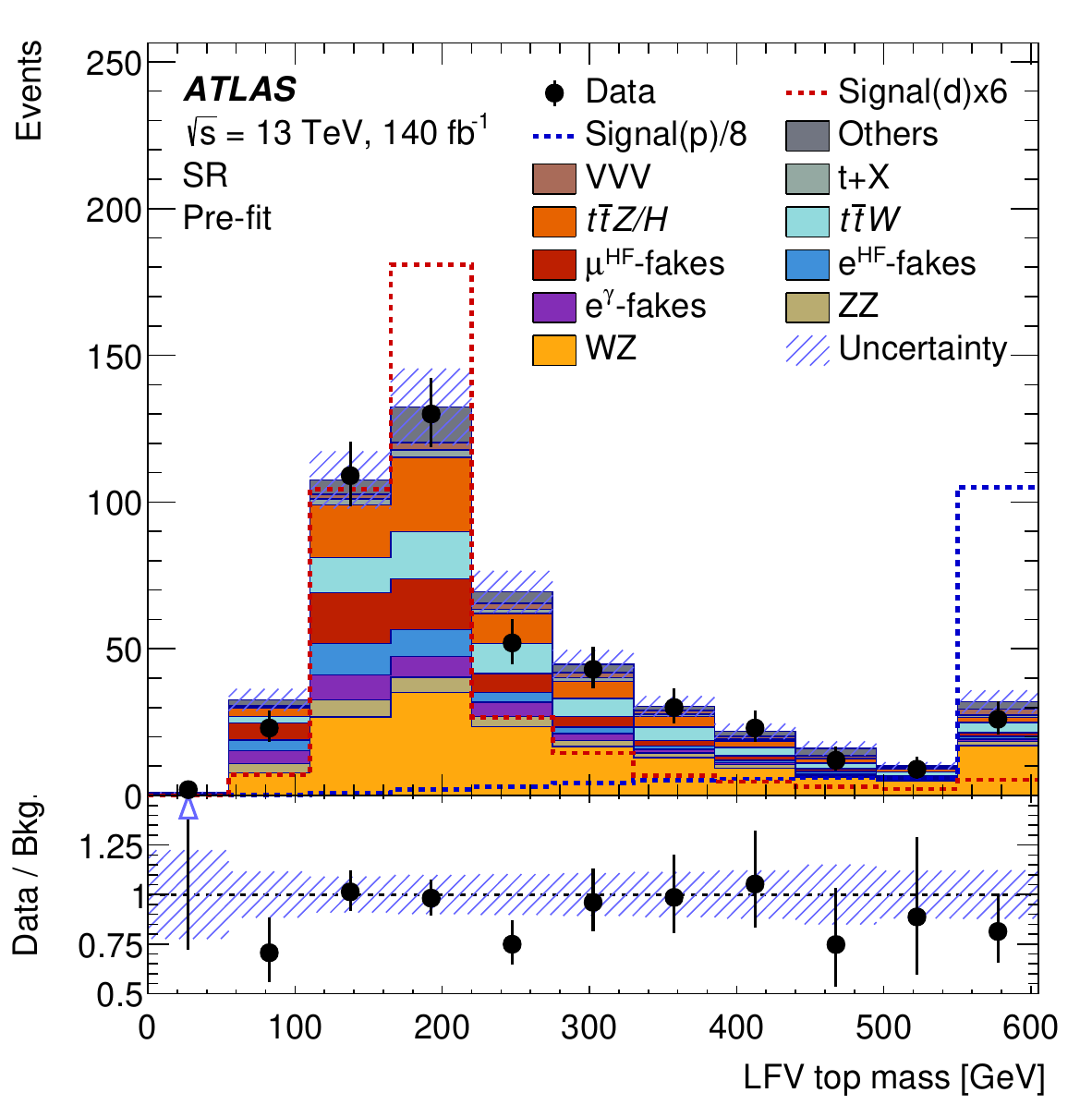}
\caption{Distribution of $m^{\mathrm{LFV}}_t$ for events with multiple jets that satisfy the common SR selection. The signal shown here has $|C^{\mathrm{tensor}}_{e\mu tu}|$ set to 0.3, and all other Wilson coefficients set to zero. `Signal (p)' and `Signal (d)' refer to the single-top production channel and $t\bar{t}$ decay channel signal, respectively.  They are individually scaled and are overlaid on the background distribution. The hatched bands indicate MC statistical and systematic uncertainties in the background predictions. The final bin includes overflow events. The lower panel shows the ratio of the data to the background prediction (`Bkg.').}
\label{fig:lfv_top_mass}
\end{figure}

For decay-channel events, the LFV top is on-shell, and so $m^{\mathrm{LFV}}_t \sim m_t$. However, for production-channel events there is no LFV top, and instead the reconstructed invariant mass is closer to that of the primary vertex, typically with $m^{\mathrm{LFV}}_t \gg m_t$. Therefore $m^{\mathrm{LFV}}_t$, shown for events satisfying the common SR selection in \cref{fig:lfv_top_mass}, provides strong separation between the decay- and production-channel signal events. These events with multiple jets are split between SR-prod and SR-decay depending on whether $m^{\mathrm{LFV}}_t$ is above or below $220$~\GeV, which was optimised using signal MC events. The full SR selection is shown in~\cref{tab:SRselection}.

\begin{table}[tb]
\centering
\caption{Event selection for the two Signal Regions.  Leptons are labelled 1, 2 or 3, ordered by decreasing \pT values.}
\begin{tabular}{l|c|cc}
\toprule
\textbf{Description} & \textbf{SR-decay} & \multicolumn{2}{c}{\textbf{SR-prod}} \\
\midrule
Number of leptons & \multicolumn{3}{c}{$N_{\text{lep}} = 3$ }\\
Lepton type & \multicolumn{3}{c}{All Tight} \\
Lepton flavours & \multicolumn{3}{c}{$2\mu 1e$, $2e 1\mu$ }\\
Third-leading lepton \pt & \multicolumn{3}{c}{$\pt^{\ell_3} > 15$~\GeV }\\
Number of $b$-tagged jets & \multicolumn{3}{c}{$N_{b-\mathrm{tags}} \leq 1$} \\
Missing transverse momentum & \multicolumn{3}{c}{$E_{\mathrm T}^{\mathrm{miss}} > 30$~\GeV} \\
$m^{\text{OSSF}}_{\ell\ell}$ closest to $Z$ mass & \multicolumn{3}{c}{$| \: m_{Z} - m_{\ell\ell}^{\mathrm{OSSF}}| > 30$~\GeV}\\
Lepton impact parameter & \multicolumn{3}{c}{$|d_0|/\sigma_{d_0} < 3 (\mu), 5(e)$} \\
\hline
Number of jets & $N_{\mathrm{jets}} \geq 2$ & \multicolumn{2}{c}{$N_{\mathrm{jets}} \geq 2$~~~~~~~~OR~~~~~~~$N_{\mathrm{jets}} =1$}  \\
Signal splitting & $m^{\mathrm{LFV}}_{t}< 220$~\GeV & $m^{\mathrm{LFV}}_{t}> 220$~\GeV\phantom{ext} & $-$\\
Background suppression & $-$ & $-$ & \phantom{ext}$H_{\mathrm T} > 400$~\GeV \\
\bottomrule
\end{tabular}
\label{tab:SRselection}
\end{table}

\subsection{Control regions}

The \CRWZ region is used to estimate the normalisation of the $WZ$ process, which is the largest prompt-lepton background entering the SRs. As with the SRs, exactly three Tight leptons are required. These must be electrons or muons, but there are no additional flavour requirements imposed. A requirement on the third-leading lepton \pT reduces the non-prompt lepton contribution, while a selection on \MET reduces the $ZZ$ background. To select a high-purity sample of $WZ$ events, an OSSF dilepton pair invariant mass within 10~\GeV of the $Z$ boson mass is required. This also ensures that \CRWZ is orthogonal to all other analysis regions.

The \CRmu and \CRel regions are used to estimate the normalisation of backgrounds with non-prompt or fake muons and electrons, respectively. These leptons stem mainly from heavy-flavour hadron decays, primarily in \ttbar and $tW$ events.
These are the largest backgrounds with non-prompt leptons that enter the SRs, and while backgrounds with non-prompt leptons from other sources are present, these are found to be small enough that dedicated control regions are not necessary. In each region, exactly two Tight leptons are required, and the third-leading lepton must be Loose. This enhances the contribution of events where the third-leading lepton is non-prompt, while the inverted transverse impact parameter requirements
ensure that \CRmu and \CRel are orthogonal to all other analysis regions. As these regions target \ttbar and $tW$ processes, up to two $b$-tagged jets are allowed, and events with an OSSF dilepton pair invariant mass within 10~\GeV of the $Z$ boson mass are vetoed.
The quantity $3p_{\mathrm{T}}^{\ell_1} + \Sigma m_{\ell \ell}^{\mathrm{OS}}$, with the sum taken over all OSSF lepton pairs and optimised in simulation, is used to suppress production-signal events that may enter these regions when a prompt lepton falls outside the acceptance~\cite{TOPQ-2023-23-custom}.
The full selection for all three CRs is shown in~\cref{tab:CRselection}.

\begin{table}[tb]
\centering
\caption{Event selection for the three Control Regions. Leptons are labelled 1, 2 or 3, according to the ordering of their $p_{\mathrm{T}}$ values.}
\begin{tabular}{l|c|c|c}
\toprule
\textbf{Description} & \textbf{\CRWZ} & \phantom{Xx}\textbf{\CRmu}\phantom{Xx} & \textbf{\CRel} \\
\midrule
Number of leptons & \multicolumn{3}{c}{$N_{\text{lep}} = 3$} \\
\midrule
Lepton types & All Tight & \multicolumn{2}{c}{$\ell_1, \ell_2$ Tight, $\ell_3$ Loose } \\
\midrule
\multirow{2}{*}{Lepton flavours} & $-$ & $\ell_1 \neq \ell_2$ &  $\ell_1 \neq \ell_2$ \\
&   & $\ell_3 = \mu$ & $\ell_3 = e$ \\
\midrule
Third-leading lepton $p_{\mathrm{T}}$ & $ > 15$~\GeV & \multicolumn{2}{c}{$-$} \\
Electron $|d_0|/\sigma_{d_0}$ & $<5$ & \multicolumn{2}{c}{$\ell_{1,2}<5, \ell_{3}\geq5$} \\
Muon $|d_0|/\sigma_{d_0}$ & $<3$ & \multicolumn{2}{c}{$\ell_{1,2}<3, \ell_{3}\geq3$} \\
$E_{\mathrm{T}}^{\mathrm{miss}}$ &$>35$~\GeV & \multicolumn{2}{c}{$-$} \\
Number of $b$-tagged jets & $\leq 1$ & \multicolumn{2}{c}{$\leq 2$} \\
$m^{\text{OSSF}}_{\ell\ell}$ closest to $Z$ mass &  $| \: m_{Z} - m_{\ell\ell}^{\mathrm{OSSF}}| < 10$~\GeV  & \multicolumn{2}{c}{$| \: m_{Z} - m_{\ell\ell}^{\mathrm{OSSF}}| > 10$~\GeV}\\
\makecell{Production signal suppression \\ $3p_{T}^{\ell_{1}}+\sum m_{\ell\ell}^{os}$ } &  $-$ & \multicolumn{2}{c}{$<400$~\GeV}\\
\midrule
\end{tabular}
\label{tab:CRselection}
\end{table}

% End of text imported from the .//sections/ObjectEventSelection.tex input file

%

%
%
%

%

% The next lines are included from the .//sections/BDT.tex input file
\section{Multivariate analysis}
\label{sec:BDT}

To obtain a better separation of the signal from the background in both the SR-decay and SR-prod regions, a multivariate technique is used, due to its capability to exploit correlations between variables in their multi-dimensional space.

This is particularly useful in SR-decay, as, unlike the production signal, the leptons in the decay signal have similar \pT values to those from the background sources, and the signal events do not exhibit strong separation in any single variable. However, the multivariate technique can also increase sensitivity significantly in SR-prod, as the characteristic final state of the production signal, with highly boosted leptons, can be fully exploited relative to the lower energy background processes.

A BDT is utilised, as implemented in the TMVA~\cite{TMVA} package embedded in ROOT~\cite{Brun:1997}. Separate BDTs are trained for SR-decay and SR-prod, including either decay- or production-signal MC samples (inclusive over all operators), and all background MC samples. The same set of 17 variables is provided to both BDTs, as listed in~\cref{tab:bdtvar}. For events with a single jet in SR-prod, where the reconstruction of the LFV top is not possible, variables such as $m_t^{\mathrm{LFV}}$ are set equal to zero. The SM top reconstruction is still performed for these events, however, to allow variables such as $m_t^{\mathrm{SM}}$ to be used for the BDT training. Similarly, for events without an OSSF or same-sign same-flavour (SSSF) lepton pair, the corresponding invariant mass is set equal to zero.

For the BDT trained in SR-decay the most important variables are $m_{\mathrm{T}}(W)$, $m_t^{\mathrm{LFV}}$, and $m_{\ell \ell}^{\mathrm{OSSF}}$, while for the BDT trained in SR-prod the most important variables are $p_{\mathrm{T}}(\ell_2)$, $m_{\mathrm{T}}(W)$, and $p_{\mathrm{T}}(e^{\mathrm{LFV}})$.
An overtraining check is performed by comparing the distributions used for training and testing using the \emph{Kolmogorov-Smirnov} test, and they are found to agree.

Following BDT training, each event in SR-decay or SR-prod is evaluated using the corresponding BDT for that region. Both regions are then binned in the BDT output score for the fit.

\begin{table}[tb]
\centering
\caption{Variables used in the multivariate analysis}
\begin{tabular}{l|l}
\toprule
\textbf{Variable} & \textbf{Label} \\
\midrule
$H_{\mathrm{T}}$ & Scalar sum of the $p_{\mathrm{T}}$ of all jets and leptons in the event \\[2pt]
$E_{\mathrm T}^{\mathrm{miss}}$ & Missing transverse momentum \\[2pt]
$N_{b\mathrm{-jets}}$ & $b$-tag multiplicity (DL1r 77\% working point) \\[2pt]
$m_{\ell \ell}^{\mathrm{OSSF}}$ & OSSF lepton pair invariant mass \\[2pt]
$m_{\ell \ell}^{\mathrm{SSSF}}$ & SSSF lepton pair invariant mass \\[2pt]
$\pT(\ell_1)$ & \pT of the leading lepton \\[2pt]
$\pT(\ell_2)$ & \pT of the sub-leading lepton \\[2pt]
$p_{\mathrm{T}}(\ell_3)$ & $p_{\mathrm T}$ of the third-leading lepton \\[2pt]
$p_{\mathrm{T}}(t^{\mathrm{LFV}})$ & $p_{\mathrm T}$ of the LFV top \\[2pt]
$m_{t}^{\mathrm{SM}}$ & SM top invariant mass \\[2pt]
${m_\mathrm{T}(W)}$ & $W$ transverse mass \\[2pt]
$m_{t}^{\mathrm{LFV}}$ & LFV top invariant mass \\[2pt]
$p_{\mathrm{T}}(e^{\mathrm{LFV}})$ & $p_{\mathrm{T}}$ of the electron associated with the LFV top decay \\[2pt]
$\pT({\mu^{\mathrm{LFV}}})$ & \pT of the muon associated with the LFV top decay \\[2pt]
$\pT(\ell^{\mathrm{SM}})$ & \pT of the lepton associated with the SM top decay \\[2pt]
$p_{\mathrm{T}}(j^{\mathrm{LFV}})$ & $p_{\mathrm{T}}$ of the jet associated with the LFV top decay \\[2pt]
$p_{\mathrm{T}}(j^{\mathrm{SM}})$ & $p_{\mathrm{T}}$ of the jet associated with the SM top decay \\
\bottomrule
\end{tabular}
\label{tab:bdtvar}
\end{table}

% End of text imported from the .//sections/BDT.tex input file

%

%

% The next lines are included from the .//sections/Systematics.tex input file
\section{Systematic uncertainties}
\label{sec:systematics}

Systematic effects may change the expected numbers of events from the signal and background processes and the shapes of the fitted distributions in the SRs and the CRs.
These effects can stem from experimental sources, including detector calibration, object reconstruction and identification efficiencies, and from the modelling of the signal and background processes.

Uncertainties due to the modelling of
the EFT signals are
estimated by considering independent variations of the renormalisation and factorisation scales by factors of 2 and 0.5, but normalising the signal to the nominal cross-section.
The uncertainties due to the PDF are calculated using the \NNPDF[3.1nlo] MC replica set. The standard deviation of this set is used to define a symmetrised uncertainty in the signal sample.
The uncertainty arising from the modelling of initial-state radiation (ISR) is evaluated by varying $\alpha_s$ in the initial-state shower using variants of the A14 tune~\cite{ATL-PHYS-PUB-2014-021}, while that arising from final-state radiation (FSR) modelling is evaluated using variations of $\mu_{r}$ for QCD emissions in the final state shower. An uncertainty due to the choice of parton shower generator is estimated by considering the change when using an alternative MC sample that uses the \HERWIG[7.1.6]~\cite{Bahr:2008pv,Bellm:2015jjp,Bellm:2017jjp} prediction with the \HERWIG[7.1] default tune~\cite{Bellm:2015jjp,Bellm:2017jjp}
instead of the \PYTHIA[8.306] generator.

Cross-section uncertainties in the SM predictions are applied to each background process, except for backgrounds whose normalisations are determined in the fit. These are
the $t\bar{t}$, $tW$, and $Z$+jets processes, which contribute to the fake or non-prompt lepton backgrounds from heavy-flavour decays in jets (denoted $e^{\mathrm{HF}}$-fake and $\mu^{\mathrm{HF}}$-fake), and the $WZ$ process.

For the $tZq$ process, a cross-section uncertainty of $14\%$ is included~\cite{TOPQ-2018-01}.
Cross-section uncertainties of $12\%$ and $10\%$, respectively, are applied to the \ttZ~and \ttH~processes, based on calculations reported in Ref.~\cite{deFlorian:2016spz}. A 30\% cross-section uncertainty is also applied to the \ttW process to account
for the uncertainty in the theoretical predictions and for the differences between the measured and predicted cross-sections for this process~\cite{Buonocore:2023ljm,Frederix:2021agh,TOPQ-2019-30,TOPQ-2020-27,CMS-TOP-21-011}. For the processes that contribute to the fake or non-prompt lepton backgrounds originating from photon conversions (labelled $e^{\gamma}$-fake), a 15\% cross-section uncertainty is applied to the $t\bar{t}\gamma$ process~\cite{TOPQ-2017-14}, and a 30\% cross-section uncertainty is applied to the $Z\gamma$ process. A 30\% cross-section uncertainty is applied to the $ZZ$ process, based on $ZZ$+jets normalisation factors determined in measurements of the $\ttbar Z$ cross-section~\cite{TOPQ-2018-08,TOPQ-2020-20},
and a 50\% cross-section uncertainty is applied to the $VVV$ process~\cite{STDM-2019-09,STDM-2020-08}.
For other processes with minor contributions to the SRs, including events with non-prompt leptons that do not originate from HF hadronic decays or photon conversions, a 50\% cross-section uncertainty is applied.

For the background processes, uncertainties due to the renormalisation and factorisation scales are estimated separately for each process, following the same procedure as for the signal.

The uncertainties due to the PDF for the \ttbar, single-top-quark, \ttW~and \ttH~processes are calculated using the \PDFforLHC[15] prescription~\cite{Butterworth:2015oua} using the Hessian sets consisting of 30 eigenvectors. Each of the 30 eigenvectors is assigned a nuisance parameter, and these nuisance parameters are correlated across all processes for which the \PDFforLHC[15] prescription is used. For the \ttZ, $t\bar{t}\gamma$, \Zjets, $Z\gamma$, and diboson background processes, the uncertainties due to the PDF are estimated following the same procedure as for the signal using the \NNPDF[3.0nlo] PDF set, separately for each sample.
For the \ttbar~process, an additional uncertainty arising from the value of $\alpha_{s}(M_{Z}) = 0.118$, used to evaluate the PDF, is estimated by using the nominal PDF set evaluated with $\alpha_{s}(M_{Z}) = 0.117$ and $\alpha_{s}(M_{Z}) = 0.119$. This uncertainty has a negligible impact and is thus neglected for the subdominant background processes.

For the \ttbar process, the size of the uncertainty due to the matching of the matrix-element calculation to the parton shower (ME--PS matching) is estimated through variation of the \PYTHIA[8] parameter $p_{\mathrm{T}}^{\mathrm{hard}}$. The nominal sample (where $p_{\mathrm{T}}^{\mathrm{hard}}=0$) is compared with a sample generated with \POWHEGBOX[v2]/\PYTHIA[8] and $p_{\mathrm{T}}^{\mathrm{hard}}=1$. To cover the choice of parton shower and hadronisation model, samples are produced with the \POWHEGBOX generator
and then showered with either \PYTHIA[8.230]
using the A14 tune
and the \NNPDF[2.3lo] PDF set, or
\HERWIG[7.1.3]~\cite{Bahr:2008pv,Bellm:2015jjp,Bellm:2017jjp},
using the \HERWIG[7.1] default set of tuned parameters.
The difference is taken as a systematic uncertainty. The uncertainty arising from the choice of the $h_{\textrm{damp}}$ parameter within \POWHEG is evaluated using an alternative \POWHEG+\Pythia[8] \ttbar sample in which $h_{\textrm{damp}} = 3.0\, m_{\mathrm{top}} $ (the default setting is $h_{\textrm{damp}} = 1.5\, m_{\mathrm{top}} $).

For the $t\bar{t}H$ and $t\bar{t}Z$ processes, the uncertainty due to the choice of parton shower and hadronisation model is evaluated using the same procedure as for the \ttbar process.

The $\ttbar W$ background has a combined uncertainty to cover the choice of generator, parton shower, and hadronisation model. This is evaluated for the QCD component as the difference between a sample generated with \SHERPA[2.2.10] and an alternative sample using \MGNLO{}+\PYTHIA[8.245] with an FxFx merging scheme~\cite{Frederix:2012ps}.
To evaluate the generator uncertainty in the ``subleading'' EW component, the \SHERPA[2.2.10] EW sample is compared with a LO \MADGRAPH{2.6.2}+\PYTHIA{8} sample that also models these EW effects.

For the \ttbar, single-top-quark, $\ttbar Z$ and $\ttbar H$ processes, the uncertainty arising from the modelling of ISR is evaluated using an independent variation of the A14 tune to its \texttt{Var3cUp} and \texttt{Var3cDown} variants~\cite{ATL-PHYS-PUB-2014-021}, corresponding to variations of $\alpha_s$ in the initial-state shower. For the \ttbar, single-top-quark and $\ttbar H$ processes the modelling of FSR is evaluated from an independent variation of the renormalisation scale for emissions from the parton shower by factors of 2.0 and 0.5.

For the $tW$ single-top-quark process, an uncertainty arising from the removal of the overlap with the \ttbar\ process is estimated by replacing the diagram-removal scheme with the diagram-subtraction scheme~\cite{Frixione:2008yi}.

The uncertainty in the combined 2015--2018 integrated luminosity is 0.83\%~\cite{DAPR-2021-01}, obtained using the LUCID-2 detector~\cite{LUCID2} for the primary luminosity measurements, complemented by measurements employing the inner detector and calorimeters.
The MC samples were reweighted to reproduce the pile-up distributions in the data separately for the three periods 2015--2016, 2017 and 2018. The uncertainty due to pile-up is evaluated by varying the correction factors used to perform the reweighting within their uncertainties.

The jet energy scale uncertainty is derived by combining information from test-beam data,
LHC collision data and simulation, and the jet energy resolution uncertainty
is obtained by combining dijet \pt-balance measurements and simulation~\cite{JETM-2018-05}.
Additional considerations related to jet flavour, pile-up corrections,
$\eta$ intercalibration and high-\pt jets are included.
A total of 31 independent contributions are considered for the jet energy scale uncertainty, and 13 for the jet energy resolution uncertainty.
There is an additional uncertainty in the efficiency of the JVT algorithm to identify and remove jets from pile-up, which is measured with $Z \to \mu^+\mu^-$ events in data using techniques similar to those used in Ref.~\cite{PERF-2014-03}.

The efficiency to correctly tag $b$-jets is measured using dileptonic \ttbar\ events~\cite{FTAG-2018-01}.
The rate for mistagging $c$-jets is measured using semileptonic \ttbar\ events,
exploiting the $c$-jets from the hadronic \Wboson-boson decays
using techniques similar to those in Ref.~\cite{FTAG-2020-08}.
The rate of mistagged light-jets is measured using a negative-tag method,
similar to that in Ref.~\cite{FTAG-2019-02}, applied to \Zjets events.
Uncertainties in the efficiency and  mistagging rates arise due to reconstructed object calibrations
and to the modelling of the different processes, and are decomposed into sets of uncorrelated sources
of uncertainty: nine, four, and four components for $b$-, $c$-, and light-jets, respectively.

Uncertainties associated with muons and electrons arise from the trigger, reconstruction,
identification and isolation efficiencies, as well as the momentum scale and resolution.
These are measured using $Z\to \ell^+\ell^-$ and $J/\psi \to \ell^+\ell^-$ events ($\ell = e,\mu$)~\cite{MUON-2022-01,EGAM-2018-01,EGAM-2021-02}.

The uncertainty originating from the limited number of simulated MC events is implemented via the Barlow--Beeston approach~\cite{Barlow:1993dm}.
Three uncertainties for each bin of the fitted distributions are considered: one for the uncertainty originating from the SM backgrounds and one for each of the cLFV production or decay signals.

% End of text imported from the .//sections/Systematics.tex input file

%

%

% The next lines are included from the .//sections/Results.tex input file
\section{Results}
\label{sec:results}

A binned profile-likelihood fit~\cite{Rolke:2004mj} to all five analysis regions is used to extract the signal strength of the cLFV process, \muCLFV, and the normalisations of the $WZ$ and HF-fake backgrounds, \kWZ, \kmuHF, and \kelHF. The SRs are binned in their BDT output discriminants for the fit, while \CReormu are binned in the third-leading lepton \pT, as  this is sensitive to the non-prompt lepton modelling. The sensitivity of the fit to the $WZ$ background, which has prompt leptons, is primarily driven by the overall yield, so \CRWZ is included in the fit with a single bin. All systematic uncertainties described in Section~\ref{sec:systematics} are included in the full likelihood as nuisance parameters (NPs), with Poisson constraints for the MC statistical uncertainties, and Gaussian constraints for all others. A pruning process is then applied to remove NPs with a small impact on \muCLFV, with the pruning threshold set at 0.3\%. The fit is evaluated with the \emph{TRExFitter} framework~\cite{TRExFitter}, which employs the \emph{RooFit} package~\cite{RooFit}. In the absence of signal, exclusion limits are set at the 95\% confidence level (CL) with the CL$_{\mathrm{S}}$ method~\cite{Read:2002hq} using MC pseudo-data ``toys''.

Separate fits are performed for each of the tensor, vector and scalar Wilson coefficients listed in \cref{tab:EFT_combination}. The Wilson coefficient in question is set to an initial value of 0.3, and the other five coefficients set to zero. The normalisation of both the decay and production signal depends on the square of the Wilson coefficient $C_i$, so the fitted value (or 95\% CL upper limit) of $|C_i|$ is extracted as

\begin{equation}
|C_i|^2 = \muCLFV \cdot 0.3^2.
\end{equation}

\begin{figure}[tb]
\centering
\subfloat[]{\includegraphics[width=0.49\textwidth]{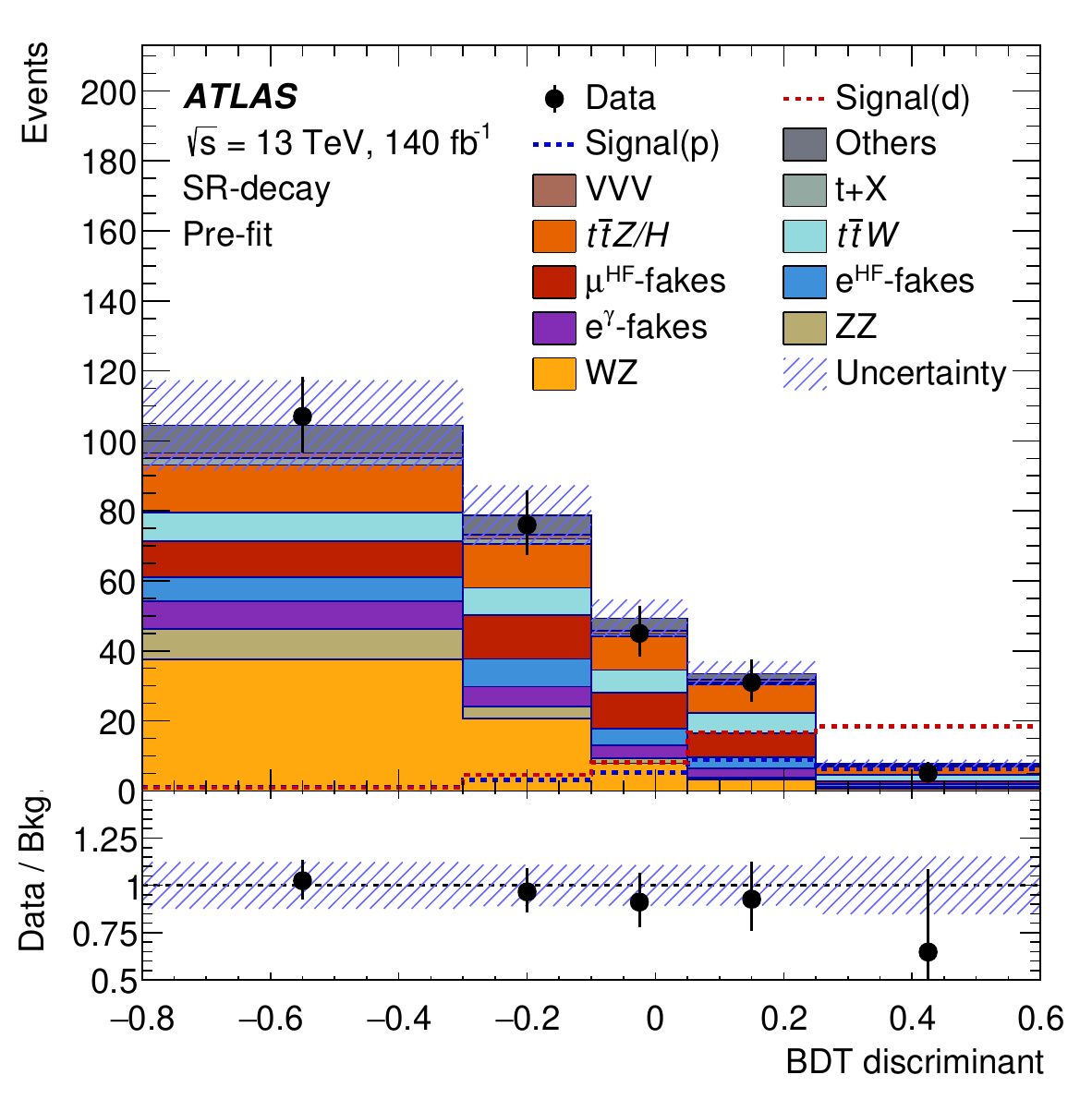}\label{fig:SR1_prefit}}
\subfloat[]{\includegraphics[width=0.49\textwidth]{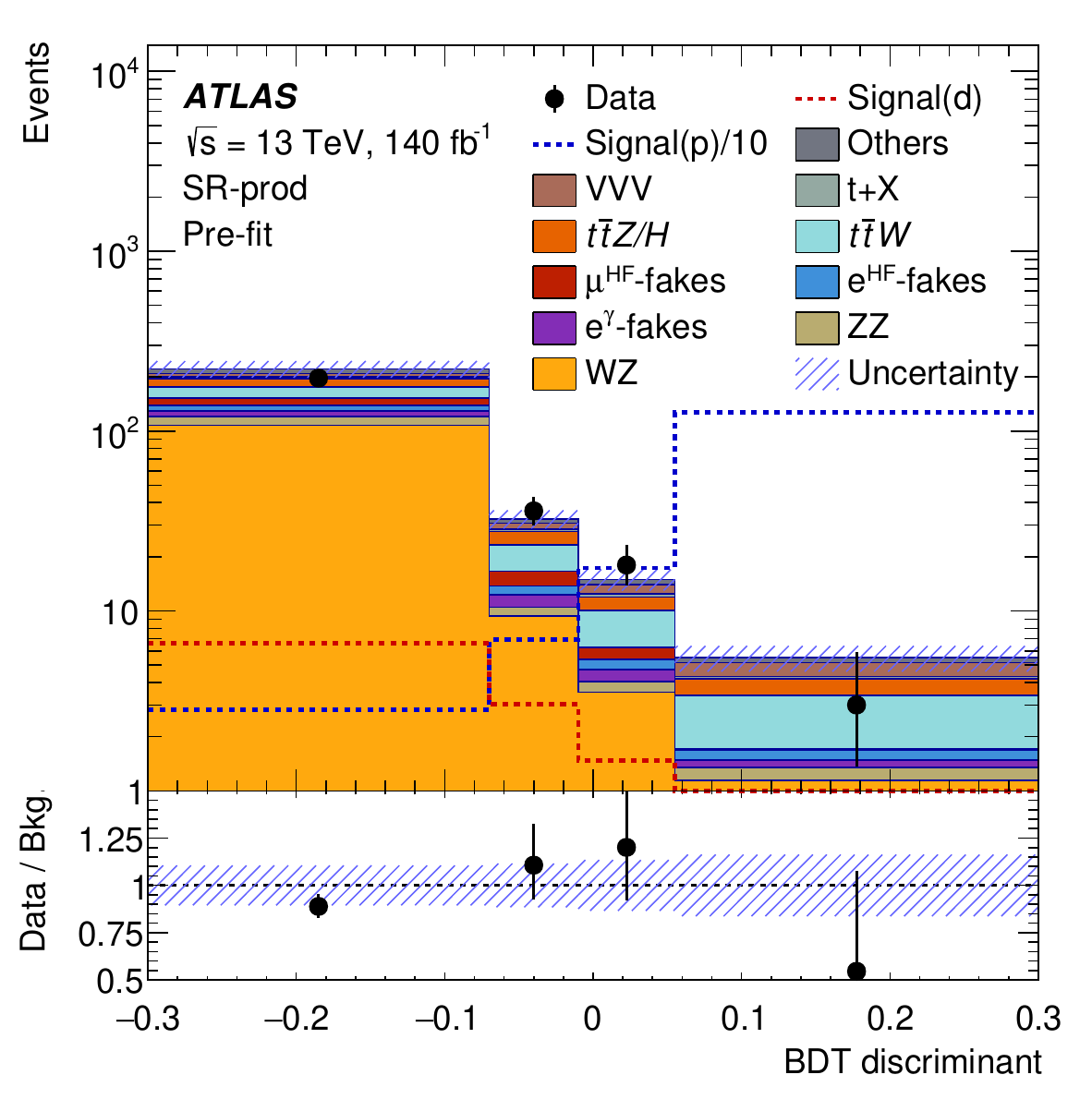}\label{fig:SR2_prefit}}

\caption{Observed event yields compared with pre-fit expectations, for the $C^{\mathrm{tensor}}_{e \mu tu}$ signal, in SR-decay~\protect\subref{fig:SR1_prefit} and SR-prod~\protect\subref{fig:SR2_prefit}. Both regions are binned in the output score of their respective BDT. `Signal (p)' and `Signal (d)' refer to the single-top production channel and \ttbar decay channel signal, respectively, and they are overlaid on the background distribution (with a scaling factor for the production signal in SR-prod). The hatched bands indicate MC statistical and systematic uncertainties in the background predictions.  The final bin of each distribution includes overflow events. The lower panel shows the ratio of the data to the background prediction (`Bkg.').}
\label{fig:prefit}
\end{figure}

Observed distributions of data in the SRs are compared with pre-fit expectations from MC simulations in~\cref{fig:prefit}. The signal shown has $|C^{\mathrm{tensor}}_{e \mu tu}| = 0.3$ and all other Wilson coefficients set to zero.

\begin{figure}[tb]
\centering
\subfloat[]{\includegraphics[width=0.49\textwidth]{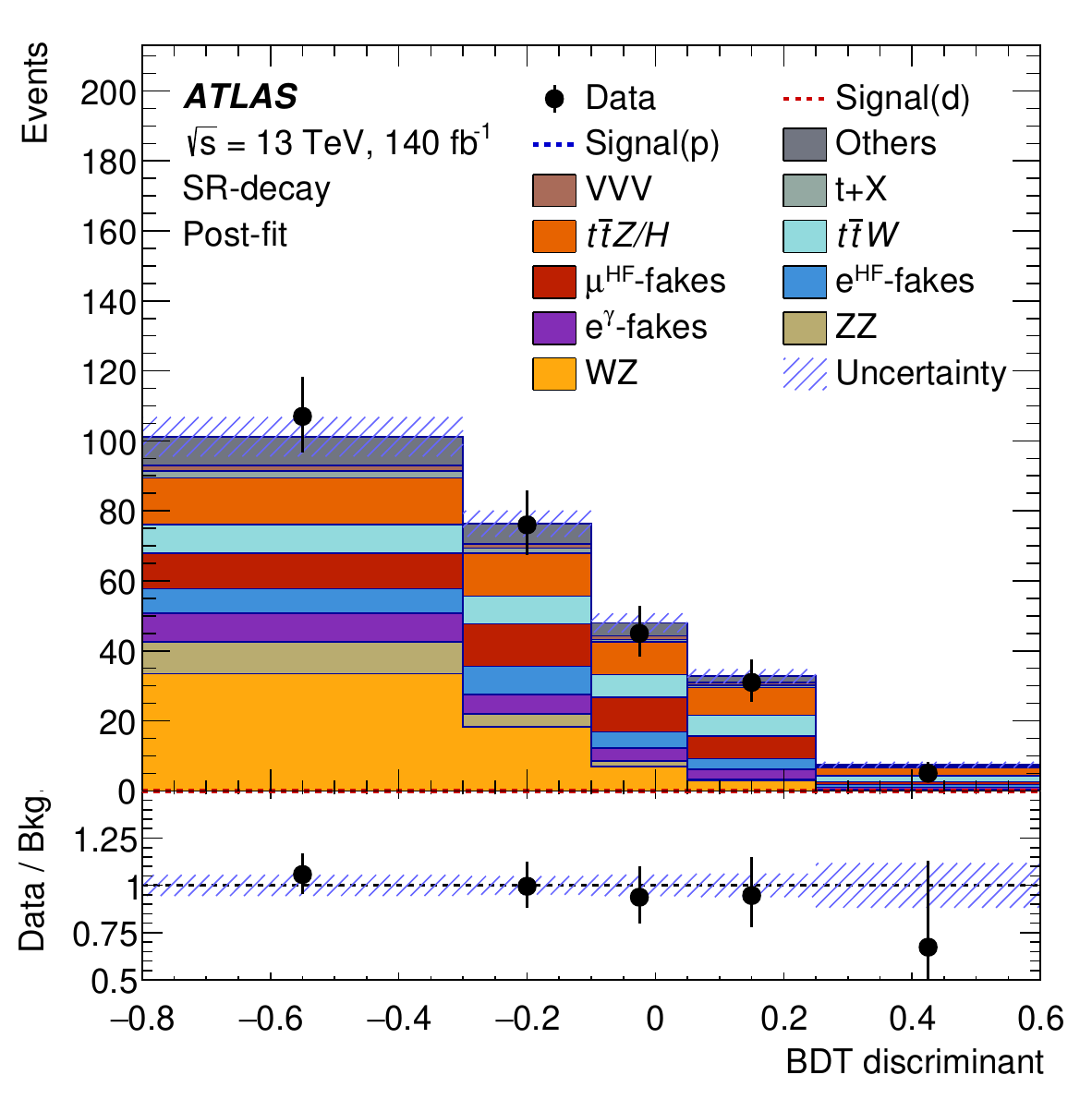}\label{fig:SR1_postfit}}
\subfloat[]{\includegraphics[width=0.49\textwidth]{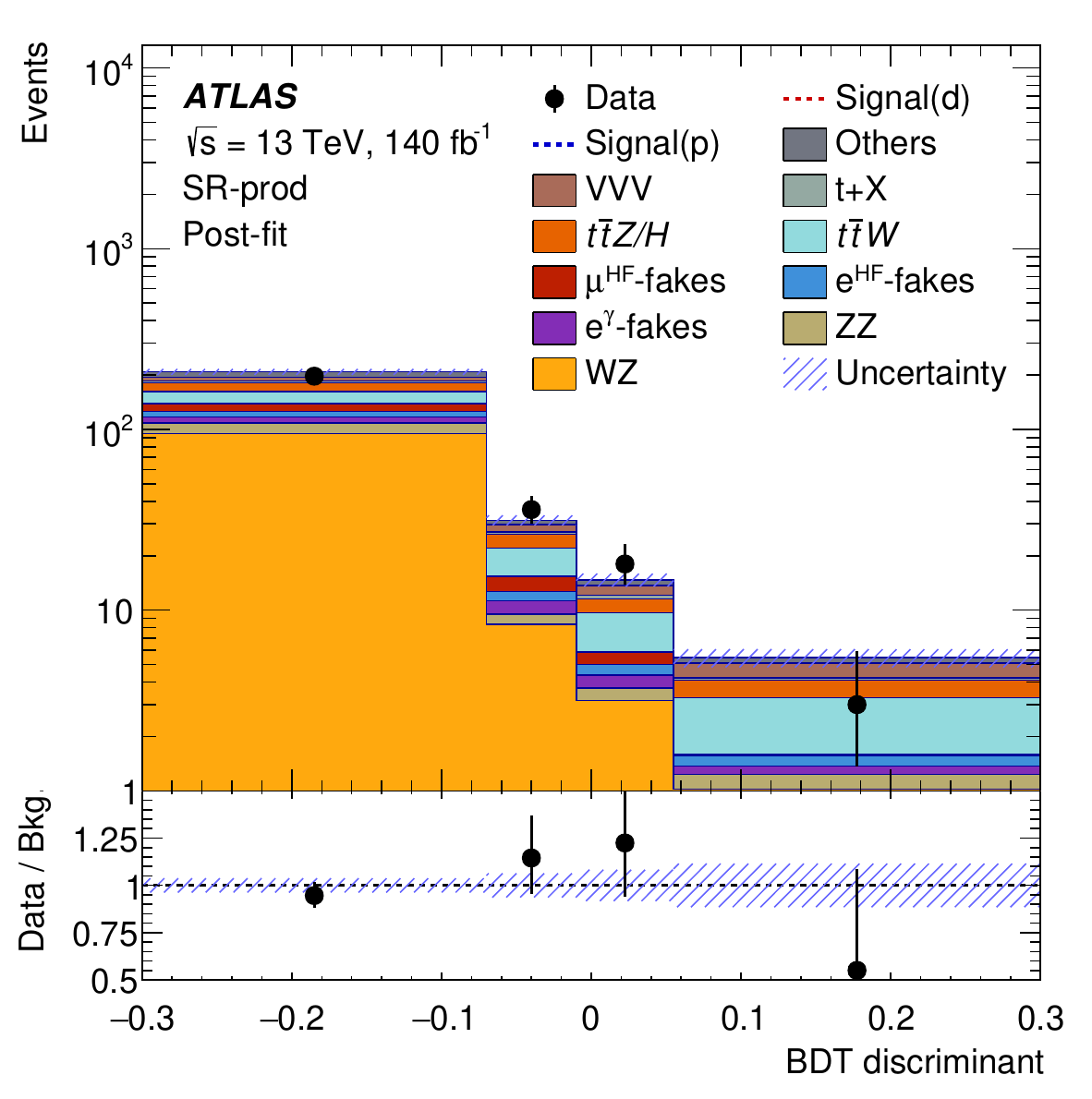}\label{fig:SR2_postfit}}

\caption{Observed event yields compared with post-fit expectations, from the fit to extract $|C^{\mathrm{tensor}}_{e \mu tu}|$, in SR-decay~\protect\subref{fig:SR1_postfit} and SR-prod~\protect\subref{fig:SR2_postfit}. Both regions are binned in the output score of their respective BDT. `Signal (p)' and `Signal (d)' refer to the single-top production channel and \ttbar decay channel signal, respectively, and they are overlaid on the background distribution (although these are not visible due to a slightly negative best-fit value of \muCLFV). The hatched bands indicate MC statistical and systematic uncertainties in the background predictions.  The final bin of each distribution includes overflow events. The lower panel shows the ratio of the data to the background prediction (`Bkg.').}
\label{fig:postfit}
\end{figure}

The corresponding post-fit distributions are shown in~\cref{fig:postfit}, taken from the fit to extract $|C^{\mathrm{tensor}}_{e \mu tu}|$. The best-fit values of the background normalisations are  $\kmuHF = 0.98^{+0.13}_{-0.11}$, $\kelHF = 1.02^{+0.12}_{-0.11}$ and $\kWZ = 0.89^{+0.12}_{-0.10}$. The HF-fake background normalisations are very close to unity, while the $WZ$ normalisation is pulled down slightly but is still consistent with unity. No significant \CLFV contributions are observed, and the data and SM prediction agree within uncertainties, with a $p$-value = 0.92 for the background-only hypothesis, calculated using the saturated model approach~\cite{Cousins:2018tiz}. Post-fit yields, determined in a fit under the background-only hypothesis, are shown in~\cref{tab:yields_postfit_BkdOnly}, along with representative pre-fit signal yields in each region.

\begin{figure}[tpb]
\centering
\subfloat[]{\label{fig:WC_limits_2D}
\includegraphics[width=0.49\textwidth]{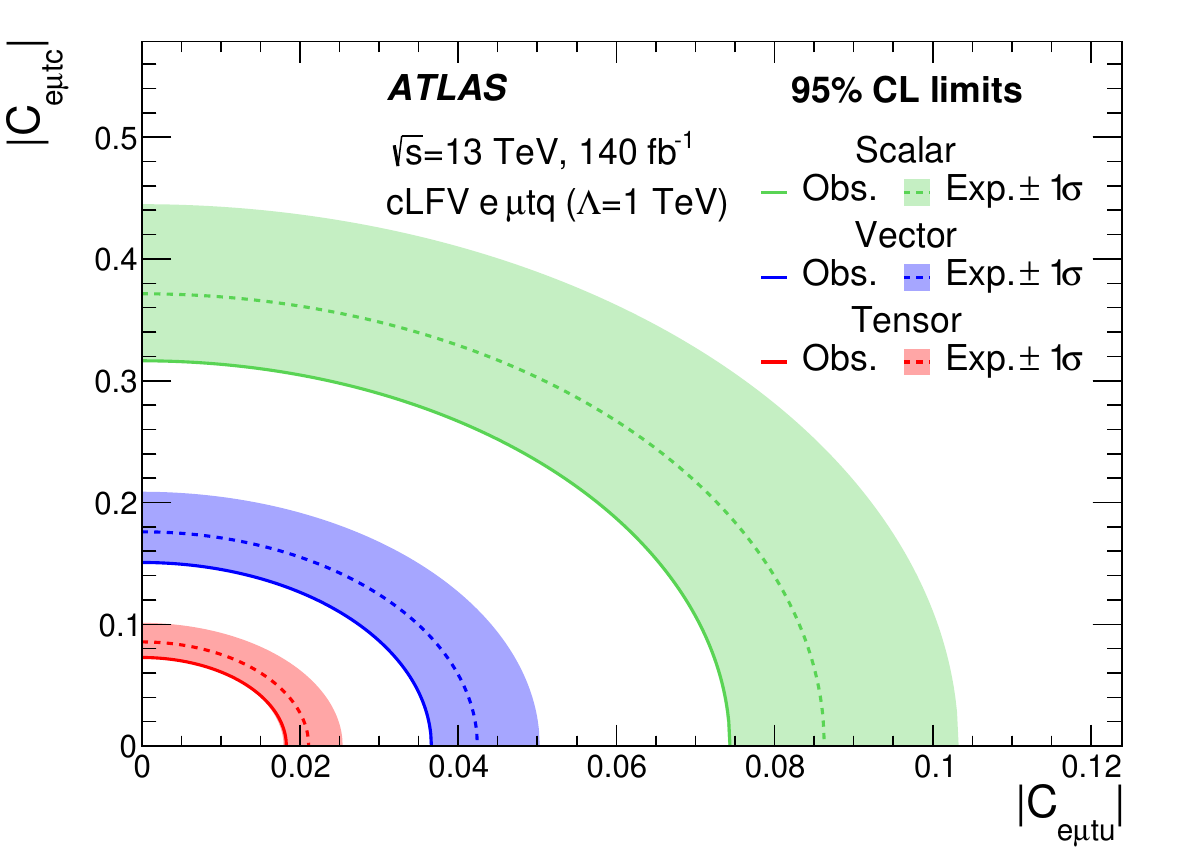}}
\subfloat[]{\label{fig:BR_limits_2D}
\includegraphics[width=0.49\textwidth]{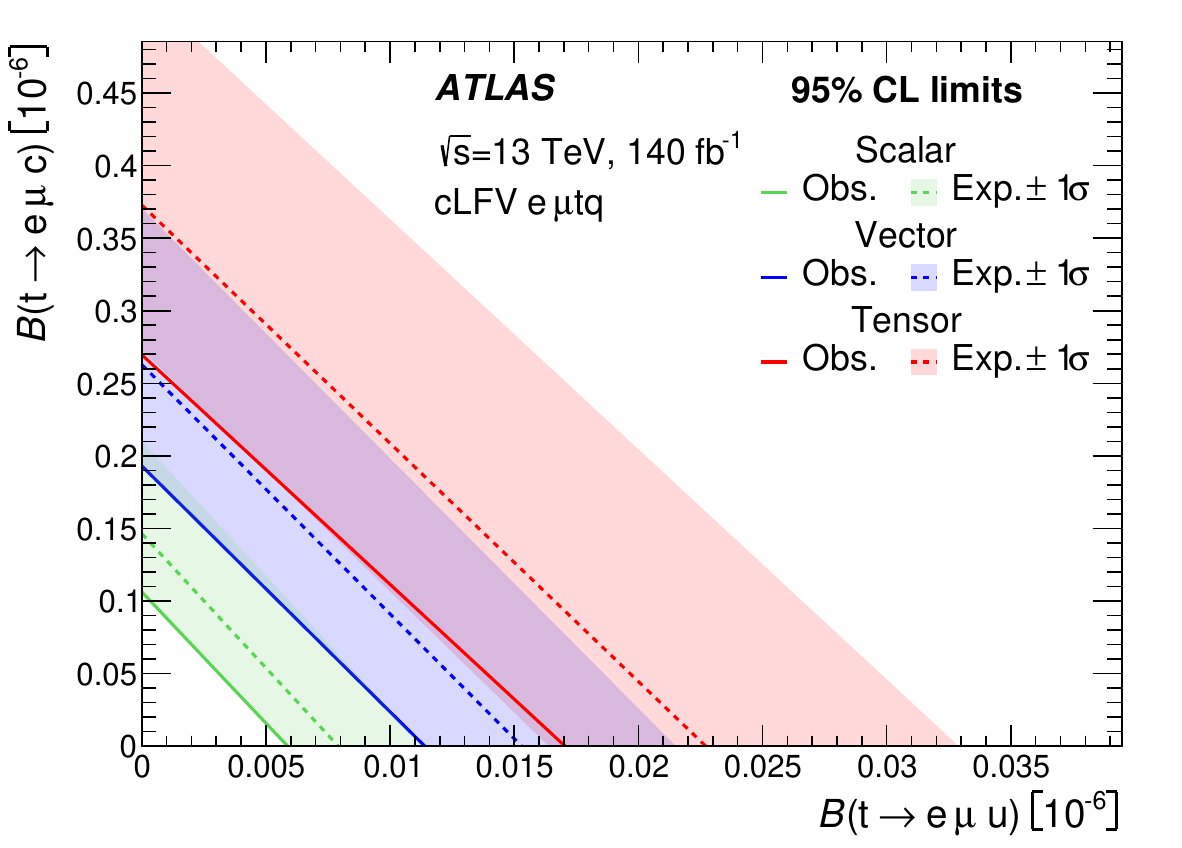}}
\caption{Observed and expected 95\% CL limits on Wilson coefficients corresponding to two-lepton two-quark EFT operators~\protect\subref{fig:WC_limits_2D}, and the branching ratios $\mathcal{B}(t\rightarrow e\mu q)$~\protect\subref{fig:BR_limits_2D}, interpolated between the $u$- and $c$-quark flavours.  The shaded areas show the one standard deviation bands around the expected limits. The interpolated lines are calculated from the individual Wilson coefficient and branching ratio limits, assuming that the signals from the $u$-quark and $c$-quark operators differ only in normalisation.}
\end{figure}

The observed and expected 95\% CL limits on the moduli of each of the six Wilson coefficients are shown in~\cref{tab:WC_limits}. The expected limits on the Wilson coefficients are converted into limits on the branching ratios $\mathcal{B}(t\rightarrow e\mu q)$ using~\cref{eq:master}. The 95\% CL limits on the branching ratios corresponding to each of the six Wilson coefficients are shown in~\cref{tab:BR_limits}. Plots of the Wilson coefficient and branching ratio limits, interpolated between the two light quark flavours, are shown in Figures \ref{fig:WC_limits_2D} and \ref{fig:BR_limits_2D}, respectively.
The data are in very good agreement with the background-only hypothesis across the range of BDT discriminant scores for both SRs. The most signal-sensitive bin for SR-prod shows a slight downward fluctuation relative to the expected background contribution (see~\cref{fig:SR2_postfit}), leading to observed limits around one sigma below the expected limits in all cases.

\begin{table}[tbp]
\centering
\caption{Post-fit event yields for each analysis region entering a fit under the background-only hypothesis, with correlations on the full systematic uncertainties taken into account. Representative pre-fit signal yields are shown in the last three rows. Signal (prod) (u/c) correspond to the single-top production channel signal, normalised to Wilson coefficient values of $|C^{\mathrm{tensor}}_{e\mu tu}| = 0.3$ or $|C^{\mathrm{tensor}}_{e\mu tc}| = 0.3$, respectively. Signal (decay) corresponds to the \ttbar decay channel signal, which has the same yield for the $u$-quark and $c$-quark operators, normalised to a Wilson coefficient value of $|C^{\mathrm{tensor}}_{e\mu tu}| = 0.3$. For all signal the scale of new physics is taken to be $\Lambda = 1$~\TeV.}
\small
\begin{tabular}{c|c|c|c|c|c}
\toprule
\textbf{Process} & \multicolumn{1}{c}{\textbf{SR-decay}} & \multicolumn{1}{c}{\textbf{SR-prod}} & \multicolumn{1}{c}{\textbf{CRWZ}} & \multicolumn{1}{c}{\textbf{CR$\mu$}} & \multicolumn{1}{c}{\textbf{CR$e$}} \\
\midrule
$WZ$   & 62 $\pm$ 5 & 107 $\pm$ 6 & 6270 $\pm$ 210 & 6.42 $\pm$ 0.28 & 3.43 $\pm$ 0.26 \\
$ZZ$   & 15 $\pm$ 5 & 15 $\pm$ 5 & 550 $\pm$ 170 & 1.8 $\pm$ 0.6 & 0.80 $\pm$ 0.25 \\
$t\bar{t}Z/H$ & 46   $\pm$ 7   & 27  $\pm$ 4   & 350  $\pm$ 60  & 2.2  $\pm$ 0.5  & 4.6 $\pm$ 1.6 \\
${t\bar{t}W}$   & 29 $\pm$ 8 & 33 $\pm$ 10 & 16 $\pm$ 4 & 1.16 $\pm$ 0.34 & 2.8 $\pm$ 0.8 \\
$\mu^{\mathrm{HF}}$-fakes & 40 $\pm$ 6 & 16.7 $\pm$ 2.2 & 96 $\pm$ 14 & 175 $\pm$ 15 & 0.51 $\pm$ 0.33 \\
$e^{\mathrm{HF}}$-fakes & 24 $\pm$ 5 & 11.0 $\pm$ 3.2 & 68 $\pm$ 34 & 0.9 $\pm$ 0.4 & 1750 $\pm$ 50 \\
$e^{\gamma}$-fakes & 20.7  $\pm$ 3.5 & 11.7 $\pm$ 2.3 & 29 $\pm$ 8 & 2.8 $\pm$ 0.4 & 19.7 $\pm$ 3.2 \\
$t+X$ & 4.9 $\pm$ 0.7 & 6.1 $\pm$ 0.9 & 340 $\pm$ 50 & 0.38 $\pm$ 0.06 & 0.51 $\pm$ 0.08 \\
$VVV$ & 4.5 $\pm$ 2.2 & 13 $\pm$ 6 & 31 $\pm$ 15 & 0.20 $\pm$ 0.10 & 0.13 $\pm$ 0.06 \\
Others & 20 $\pm$ 8    & 17 $\pm$ 7    & 140 $\pm$ 60  & 9.2 $\pm$ 3.3 & 68 $\pm$ 30 \\
\midrule
Total & 265 $\pm$ 12 & 258 $\pm$ 10 & 7900 $\pm$ 90 & 200 $\pm$ 14 & 1850 $\pm$ 40 \\
\midrule
Data & 264  & 254 & 7906  & 202  & 1855 \\
\midrule
\midrule
Signal (decay) & 48.7 $\pm$ 2.5 & 11.8 $\pm$ 1.7 & 5.6 $\pm$ 0.6 & 0.53 $\pm$ 0.13 & 0.42 $\pm$ 0.09 \\
Signal (prod) (u)   & 24.5 $\pm$ 3.4 & 1540 $\pm$ 170 & 25 $\pm$ 4 & 0.13 $\pm$ 0.10 & 0.005 $\pm$ 0.022 \\
Signal (prod) (c)   & 3.6 $\pm$ 0.7 & 104 $\pm$ 12 & 2.9 $\pm$ 0.4 & 0.039 $\pm$ 0.029 & 0.021 $\pm$ 0.016 \\
\bottomrule
\end{tabular}
\label{tab:yields_postfit_BkdOnly}
\end{table}

\begin{table}[tbp]
\begin{center}
\caption{Observed and expected 95\% CL upper limits on the moduli of Wilson coefficients corresponding to the 2Q2L EFT operators outlined in \cref{tab:EFT_combination}, which introduce \CLFV top production and decay in the \emu channel. The scale of new physics is taken to be $\Lambda = 1$~\TeV.}
\begin{tabular}{c|c|c|c|c}
\toprule
\makecell{\textbf{cLFV} \\ \textbf{coupling}} & \textbf{Lorentz Structure} & \makecell{\textbf{Wilson} \\ \textbf{coefficient}} & \makecell{\textbf{Expected 95\% CL limits} \\ \textbf{($\pm ~ 1\sigma$ range)}} & \makecell{\textbf{Observed} \\ \textbf{95\% CL limits}} \\
\midrule
$e\mu tu$ & Tensor & $|C^{\mathrm{tensor}}_{e\mu tu}|$ & 0.021 (0.018 -- 0.025) & 0.018 \\[2pt]
& Vector &$|C^{\mathrm{vector}}_{e\mu tu}|$ & 0.042 (0.036 -- 0.050) & 0.037 \\[2pt]
& Scalar &$|C^{\mathrm{scalar}}_{e\mu tu}|$ & 0.086 (0.074 -- 0.103) & 0.074 \\[2pt]
\midrule
$e\mu tc$ & Tensor &$|C^{\mathrm{tensor}}_{e\mu tc}|$ & 0.086 (0.073 -- 0.101) & 0.073 \\[2pt]
& Vector &$|C^{\mathrm{vector}}_{e\mu tc}|$
& 0.18 (0.15 -- 0.21) & 0.15 \\[2pt]
& Scalar &$|C^{\mathrm{scalar}}_{e\mu tc}|$
& 0.37 (0.32 -- 0.43) & 0.32 \\
\bottomrule
\end{tabular}
\label{tab:WC_limits}
\end{center}
\end{table}

\begin{table}[tbp]
\begin{center}
\caption{Observed and expected 95\% CL upper limits on \CLFV branching ratios $\mathcal{B}(t\rightarrow e\mu q)$, calculated from the Wilson coefficients corresponding to the 2Q2L EFT operators outlined in \cref{tab:EFT_combination}, which introduce \CLFV top production and decay in the \emu channel.}
\begin{tabular}{c|c|c|c|c}
\toprule
\makecell{\textbf{cLFV} \\ \textbf{coupling}} & \textbf{Lorentz Structure} & \makecell{\textbf{Branching} \\ \textbf{ratio}} & \makecell{\textbf{Expected 95\% CL limits} \\ \textbf{$\left[10^{-6}\right]$ ($\pm~1\sigma$ range)}} & \makecell{\textbf{Observed} \\ \textbf{95\% CL limits $\left[10^{-6}\right]$}} \\
\midrule
$e\mu tu$ & Tensor & $\mathcal{B}(t\rightarrow e\mu u)$ & 0.023 (0.016 -- 0.033) & 0.017 \\[2pt]
& Vector & & 0.015 (0.011 -- 0.022) & 0.011 \\[2pt]
& Scalar & & 0.008 (0.006 -- 0.011) & 0.006 \\[2pt]
\midrule
$e\mu tc$ & Tensor &$\mathcal{B}(t\rightarrow e\mu c)$
& 0.37 (0.27 -- 0.52) & 0.27 \\[2pt]
& Vector &
& 0.26 (0.20 -- 0.37) & 0.19 \\[2pt]
& Scalar &
& 0.15 (0.11 -- 0.21) & 0.11 \\
\bottomrule

\end{tabular}
\label{tab:BR_limits}
\end{center}
\end{table}

The sensitivity for all operators is dominated by the single-top production channel signal, due to its much larger cross-section compared with the $t\bar{t}$ decay channel (see Table \ref{tab:clfv_xsecs_all}). For the $u$-quark operators the signal-strength sensitivity of the decay channel alone is about 1\% of that obtained when both channels are combined, whereas for the $c$-quark operators it is about 20\% of that obtained in the combined analysis. This is due to the relatively lower cross-section for the $c$-type production signal, which requires a $c$-quark to be present in the initial state.

All limits are dominated by the statistical uncertainty; removing all systematic uncertainties only improves the limits by up to 2\%. The largest sources of systematic uncertainty relate to the modelling of the $VVV$, $t\bar{t}W$, and $t\bar{t}Z$ processes, which are present in the final bin of the BDT discriminant distribution in SR-prod. The sensitivity to the signal process is primarily driven by this bin, and so the systematic uncertainties that most affect the expected background yield will have the largest impact. Another source of systematic uncertainty stems from the modelling of the production signal process. The largest single effect on the signal yield is around $10\%$, which is due to the choice of parton shower generator.

The results presented here can be compared with those determined by the CMS Collaboration for the same trileptonic channel involving electrons and muons~\cite{CMS-TOP-22-005}. The expected limits on \CLFV production in the two cases are very similar, giving almost identical limits on both the Wilson coefficients and the branching ratios. However, the ATLAS data show a small downward fluctuation relative to the expected background contribution in the most signal-sensitive region, leading to more stringent observed limits.

\FloatBarrier

% End of text imported from the .//sections/Results.tex input file

%
%

%
%
%
%
%
%

%
%
\FloatBarrier

%

% The next lines are included from the .//sections/Conclusion.tex input file
\section{Conclusions}

A search for charged lepton-flavour violation in top-quark events is presented, using data collected by the ATLAS detector in proton--proton collisions at the LHC at a centre-of-mass energy of $\sqrt{s} = 13$~\TeV, corresponding to an integrated luminosity of $140$~\ifb{}.  The decay and production processes, $\ttbar \to (\ell\ell^{'}q)(W(\to \ell\nu) b)$ and $qg \to t\ell\ell^{'}$ respectively, where $\ell^{(')}=e,\mu$,
were considered, whereby the SM top produces a single charged lepton ($e$ or $\mu$) in the final state.
The decay (production) target topology is thus characterised by the presence of three isolated charged leptons, at least two (one) jets, and missing transverse momentum from the undetected neutrino. A multivariate technique is employed to separate potential signal from the SM background, and is optimised separately for the decay and production processes.

No evidence for a signal is seen, and the SM prediction is found to agree with the data. Upper limits are placed at 95\% CL on the Wilson coefficients corresponding to scalar, vector, and tensor contributions to the \CLFV $e\mu tq$ interaction. These range from $|C^{\mathrm{tensor}}_{e\mu tu}|/\Lambda^2 < 0.018~\TeV^{-2}$ to $|C^{\mathrm{scalar}}_{e\mu tc}|/\Lambda^2 < 0.32~\TeV^{-2}$ for the tensor $u$-quark and scalar $c$-quark couplings, respectively. These Wilson coefficient limits are translated into upper limits on \CLFV top decay branching ratios, ranging from $\mathcal{B}(t \to e\mu u) < 6 \cdot 10^{-9}$ to $\mathcal{B}(t \to e\mu c) < 2.7 \cdot 10^{-7}$ for the scalar $u$-quark and tensor $c$-quark couplings, respectively.
The sensitivity for all operators is dominated by the contribution of the single-top production channel.  These are the most stringent limits to date on the $e\mu tq$ process.

\clearpage

% End of text imported from the .//sections/Conclusion.tex input file

%

\section*{Acknowledgements}
%

%

% The next lines are included from the .//acknowledgements/Acknowledgements.tex input file
%
%

%
%

We thank CERN for the very successful operation of the LHC and its injectors, as well as the support staff at
CERN and at our institutions worldwide without whom ATLAS could not be operated efficiently.

The crucial computing support from all WLCG partners is acknowledged gratefully, in particular from CERN, the ATLAS Tier-1 facilities at TRIUMF/SFU (Canada), NDGF (Denmark, Norway, Sweden), CC-IN2P3 (France), KIT/GridKA (Germany), INFN-CNAF (Italy), NL-T1 (Netherlands), PIC (Spain), RAL (UK) and BNL (USA), the Tier-2 facilities worldwide and large non-WLCG resource providers. Major contributors of computing resources are listed in Ref.~\cite{ATL-SOFT-PUB-2026-001}.

We gratefully acknowledge the support of ANPCyT, Argentina; YerPhI, Armenia; ARC, Australia; BMWFW and FWF, Austria; ANAS, Azerbaijan; CNPq and FAPESP, Brazil; NSERC, NRC and CFI, Canada; CERN; ANID, Chile; CAS, MOST, NSFC and FRFCU, China; Minciencias, Colombia; MEYS CR, Czech Republic; DNRF and DNSRC, Denmark; IN2P3-CNRS and CEA-DRF/IRFU, France; SRNSFG, Georgia; BMFTR, HGF and MPG, Germany; GSRI, Greece; RGC and Hong Kong SAR, China; ICHEP and Academy of Sciences and Humanities, Israel; INFN, Italy; MEXT and JSPS, Japan; CNRST, Morocco; NWO, Netherlands; RCN, Norway; MNiSW, Poland; FCT, Portugal; MNE/IFA, Romania; MSTDI, Serbia; MSSR, Slovakia; ARIS and MVZI, Slovenia; DSI/NRF, South Africa; MICIU/AEI, Spain; SRC and Wallenberg Foundation, Sweden; SERI, SNSF and Cantons of Bern and Geneva, Switzerland; NSTC, Taipei; TENMAK, T\"urkiye; STFC/UKRI, United Kingdom; DOE and NSF, United States of America.

Individual groups and members have received support from BCKDF, CANARIE, CRC and DRAC, Canada; CERN-CZ and PRIMUS, Czech Republic; COST, ERC, ERDF, Horizon 2020 and Marie Sk{\l}odowska-Curie Actions, European Union; Investissements d'Avenir Labex, Investissements d'Avenir Idex and ANR, France; DFG and AvH Foundation, Germany; Herakleitos, Thales and Aristeia programmes co-financed by EU-ESF and the Greek NSRF, Greece; BSF-NSF and MINERVA, Israel; NCN and NAWA, Poland; La Caixa Banking Foundation, CERCA and AGAUR programs from Generalitat de Catalunya and PROMETEO and GenT Programmes Generalitat Valenciana, Spain; G\"{o}ran Gustafssons Stiftelse, Sweden; The Royal Society and Leverhulme Trust, United Kingdom; Eric and Wendy Schmidt Fund for Strategic Innovation, United States of America.

In addition, individual members wish to acknowledge support from Chile: Agencia Nacional de Investigaci\'on y Desarrollo (ANID FONDECYT reg. 1230987, FONDECYT 1230812, FONDECYT 1240864, Fondecyt 3240661, Fondecyt Regular 1240721); China: Fundamental Research Funds for the Central Universities (010-63263105), Chinese Ministry of Science and Technology (MOST-2023YFA1605700, MOST-2023YFA1609300), National Natural Science Foundation of China (NSFC 12275265, NSFC-W2543005); Czech Republic: Czech Science Foundation (GACR - 24-11373S), Ministry of Education Youth and Sports (ERC-CZ-LL2327, FORTE CZ.02.01.01/00/22\_008/0004632); EU: H2020 European Research Council (ERC - 101002463); European Union: European Research Council (BARD No. 101116429, ERC - 101219398, ERC 101089007), European Regional Development Fund (HE COFUND GA No.101081355, ERDF), Marie Sklodowska-Curie Actions (GAP-101168829); France: Agence Nationale de la Recherche (ANR-22-EDIR-0002, ANR-24-CE31-0504-01); China: Research Grants Council (GRF); Italy: Istituto Nazionale di Fisica Nucleare (LHC-MIUR - 28003/2025), Ministero dell'Università e della Ricerca (NextGenEU I53D23000820006 M4C2.1.1, SOE2024\_0000023); Japan: Japan Society for the Promotion of Science (JSPS KAKENHI  JP25H0063, JSPS KAKENHI JP22H04944, JSPS KAKENHI JP24K23939, JSPS KAKENHI JP24KK0251, JSPS KAKENHI JP25H00650, JSPS KAKENHI JP25H01291, JSPS KAKENHI JP25K01011, JSPS KAKENHI JP25K01023, JSPS KAKENHI JP25KK0047); Poland: Polish National Science Centre (NCN OPUS 2023/51/B/ST2/02507, NCN OPUS nr 2022/47/B/ST2/03059, UMO-2019/34/E/ST2/00393, UMO-2021/42/E/ST2/00350, UMO-2022/47/O/ST2/00148, UMO-2023/49/B/ST2/04085, UMO-2023/51/B/ST2/00920, UMO-2024/53/N/ST2/00869); Spain: Agència de Gestió d'Ajuts Universitaris i de Recerca. (AGAUR - 2023 BP 00141), Ministry of Science and Innovation (RYC2019-028510-I, RYC2020-030254-I, RYC2021-031273-I, RYC2022-038164-I), Ministerio de Ciencia, Innovación y Universidades/Agencia Estatal de Investigaci\'on (PID2022-142604OB-C22); Sweden: Carl Trygger Foundation (Carl Trygger Foundation CTS 22:2312), Swedish Research Council (Swedish Research Council 2023-04654, VR 2021-03651, VR 2022-03845, VR 2022-04683, VR 2023-03403, VR 2024-05451, VR 2025-05940), Knut and Alice Wallenberg Foundation (KAW 2023.0366); United Kingdom: The Binks Trust, Royal Society (NIF-R1-231091); United States of America: U.S. Department of Energy (ECA DE-AC02-76SF00515), John Templeton Foundation (John Templeton Foundation 63206), Neubauer Family Foundation.

%
%

% End of text imported from the .//acknowledgements/Acknowledgements.tex input file

%
%
%
%
%
%
%
%
%

%
\clearpage
\printbibliography
\clearpage
\input{atlas_authlist}
%
%
%

%

%
%
%
%

\end{document}

%% file: atlas_authlist.tex
% ATLAS Collaboration author list
% Reference date of TOPQ-2023-35 is 2026-08-28
% Author list last updated on date 01-OCT-26
% Data extracted on 01-Oct-2026 for paper reference TOPQ-2023-35
% at 9:18am
 
\begin{flushleft}
\hypersetup{urlcolor=black}
{\Large The ATLAS Collaboration}

\bigskip

\AtlasOrcid[0000-0002-6665-4934]{G.~Aad}$^\textrm{\scriptsize 102}$,
\AtlasOrcid[0000-0001-7616-1554]{E.~Aakvaag}$^\textrm{\scriptsize 17}$,
\AtlasOrcid[0000-0002-5888-2734]{B.~Abbott}$^\textrm{\scriptsize 121}$,
\AtlasOrcid[0000-0002-0287-5869]{S.~Abdelhameed}$^\textrm{\scriptsize 83b}$,
\AtlasOrcid[0000-0002-1002-1652]{K.~Abeling}$^\textrm{\scriptsize 54}$,
\AtlasOrcid[0000-0001-5763-2760]{N.J.~Abicht}$^\textrm{\scriptsize 48}$,
\AtlasOrcid[0000-0002-8496-9294]{S.H.~Abidi}$^\textrm{\scriptsize 30}$,
\AtlasOrcid[0009-0003-6578-220X]{M.~Aboelela}$^\textrm{\scriptsize 44}$,
\AtlasOrcid[0000-0002-9987-2292]{A.~Aboulhorma}$^\textrm{\scriptsize 36e}$,
\AtlasOrcid[0000-0001-5329-6640]{H.~Abramowicz}$^\textrm{\scriptsize 154}$
% ...
\bigskip
\\

% ...
$^{17}$Department for Physics and Technology, University of Bergen, Bergen; Norway.\\
% ...
$^{30}$Physics Department, Brookhaven National Laboratory, Upton NY; United States of America.\\
% ...
$^{36}$$^{(a)}$Facult\'e des Sciences Ain Chock, Universit\'e Hassan II de Casablanca;$^{(b)}$Facult\'{e} des Sciences, Universit\'{e} Ibn-Tofail, K\'{e}nitra;$^{(c)}$Facult\'e des Sciences Semlalia, Universit\'e Cadi Ayyad, LPHEA-Marrakech;$^{(d)}$LPMR, Facult\'e des Sciences, Universit\'e Mohamed Premier, Oujda;$^{(e)}$Facult\'e des sciences, Universit\'e Mohammed V, Rabat;$^{(f)}$Institute of Applied Physics, Mohammed VI Polytechnic University, Ben Guerir; Morocco.\\
% ...
$^{44}$Physics Department, Southern Methodist University, Dallas TX; United States of America.\\
% ...
$^{48}$Fakult\"{a}t Physik, Technische Universit{\"a}t Dortmund, Dortmund; Germany.\\
% ...
$^{54}$II. Physikalisches Institut, Georg-August-Universit\"{a}t G\"ottingen, G\"ottingen; Germany.\\
% ...
$^{83}$$^{(a)}$Khalifa University of Science and Technology, Abu Dhabi;$^{(b)}$New York University Abu Dhabi, Abu Dhabi;$^{(c)}$United Arab Emirates University, Al Ain;$^{(d)}$University of Sharjah, Sharjah; United Arab Emirates.\\
% ...
$^{102}$CPPM, Aix-Marseille Universit\'e, CNRS/IN2P3, Marseille; France.\\
% ...
$^{121}$Homer L. Dodge Department of Physics and Astronomy, University of Oklahoma, Norman OK; United States of America.\\
% ...
$^{154}$Raymond and Beverly Sackler School of Physics and Astronomy, Tel Aviv University, Tel Aviv; Israel.\\
% ...
$^{a}$ Also at Affiliated with an institute formerly covered by a cooperation agreement with CERN.\\
$^{b}$ Also at An-Najah National University, Nablus; Palestine.\\
$^{c}$ Also at Borough of Manhattan Community College, City University of New York, New York NY; United States of America.\\
$^{d}$ Also at Center for Interdisciplinary Research and Innovation (CIRI-AUTH), Thessaloniki; Greece.\\
$^{e}$ Also at Centre of Physics of the Universities of Minho and Porto (CF-UM-UP); Portugal.\\
$^{f}$ Also at CERN, Geneva; Switzerland.\\
% ...

\end{flushleft}

% Created with Glance <Atlas.Glance@cern.ch>